\documentclass[12pt]{article}
\usepackage{amsmath,amssymb}
\usepackage{bm}
\usepackage{color}
\usepackage{graphicx}
\usepackage{cite}
\usepackage{mathtools}

\usepackage{slashed}

\usepackage{here}

\begin{document}

\title{
\begin{flushright}
\ \\*[-80pt]
\begin{minipage}{0.22\linewidth}
\normalsize
%arXiv:YYMM.NNNN \\
EPHOU-22-020\\
KYUSHU-HET-250\\
KOBE-TH-22-06\\*[50pt]
\end{minipage}
\end{flushright}
% Title
{\Large \bf
Zero-mode wave functions by localized gauge fluxes
\\*[20pt]}}
% /Title

\author{
Tatsuo Kobayashi $^{1}$,
%\footnote{A's mail}
~Hajime Otsuka $^{2}$,
%\footnote{B's mail}
~Makoto Sakamoto $^{3}$, \\
%\footnote{C's mail}
~Maki Takeuchi $^{3}$,
%\footnote{D's mail}
Yoshiyuki Tatsuta $^{4}$,
%\footnote{E's mail}
and
~Hikaru Uchida $^{1}$
%\footnote{F's mail}
\\*[20pt]
\centerline{
\begin{minipage}{\linewidth}
\begin{center}
$^1${\it \normalsize
Department of Physics, Hokkaido University, Sapporo 060-0810, Japan} \\*[5pt]
$^2${\it \normalsize
Department of Physics, Kyushu University, 744 Motooka, Nishi-ku, Fukuoka 819-0395, Japan} \\*[5pt]
$^3${\it \normalsize
Department of Physics, Kobe University, Kobe 675-8501, Japan} \\*[5pt]
$^4${\it \normalsize
Scuola Normale Superiore and INFN, Piazza dei Cavalieri 7, 56126 Pisa, Italy} \\*[5pt]
\end{center}
\end{minipage}}
\\*[50pt]}

\date{
\centerline{\small \bf Abstract}
\begin{minipage}{0.9\linewidth}
\medskip
\medskip
\small
We study chiral zero-mode wave functions on blow-up manifolds of $T^2/Z_N$ orbifolds with both bulk and localized magnetic flux backgrounds. 
We introduce a singular gauge transformation in order to remove $Z_N$ phases for $Z_N$ twisted boundary condition of matter fields.
We compute wave functions of not only bulk zero modes but also localized modes at the orbifold singular points, which correspond to new zero modes induced by localized flux.
By studying their Yukawa couplings, it turns out that only three patterns of Yukawa couplings are allowed.
Our theory has a specific coupling selection rule.
\end{minipage}
}

\begin{titlepage}
\maketitle
\thispagestyle{empty}
\end{titlepage}

\newpage

\tableofcontents

% ------------------------------------------------------ %
% ------------------------------------------------------ %
% ------------------------------------------------------ %
% ------------------------------------------------------ %

\section{Introduction}
\label{sec:intro}

A higher dimensional theory such as the string theory is one of candidates for underlying theory beyond the Standard Model.
In such a theory, how to compactify extra dimensions is important to derive a four-dimensional (4D) low-energy effective field 
theory.
Among other compactifications, toroidal orbifold compactification~\cite{Dixon:1985jw,Dixon:1986jc} is quite simple, but attractive, because 
in principle, one can calculate all couplings of 4D effective theory.

In particular, torus and its orbifold models with background magnetic fluxes lead to a 4D chiral theory, 
where zero-modes are degenerate, and their number corresponds to the generation number.
The number of zero-modes in magnetized torus and orbifold models is 
determined by the size of magnetic flux~\cite{Cremades:2004wa,Abe:2008fi,Abe:2013bca}.
Indeed, three generations models were classified \cite{Abe:2008sx,Abe:2015yva}.
The index theorem was also studied in magnetized orbifold models \cite{Sakamoto:2020pev,Sakamoto:2020vdy}.
Furthermore, Yukawa couplings~\cite{Cremades:2004wa} as well as higher dimensional couplings~\cite{Abe:2009dr} were  calculated. 
Then, realistic quark and lepton masses and their mixing angle as well as the CP phase were derived 
\cite{Abe:2012fj,Abe:2014vza,Fujimoto:2016zjs,Kobayashi:2016qag,Kikuchi:2021yog,Kikuchi:2022geu,Hoshiya:2022qvr}.
Recently, it was found that the flavor structure in 4D effective theory is controlled by 
the modular symmetry \cite{Kobayashi:2018rad,Kobayashi:2018bff,Kariyazono:2019ehj,Ohki:2020bpo,Kikuchi:2020frp,Kikuchi:2020nxn,
Kikuchi:2021ogn,Almumin:2021fbk,Tatsuta:2021deu,Kikuchi:2022bkn}.

Toroidal orbifolds are quite special from the viewpoint of Calabi-Yau manifolds and their moduli spaces.
Indeed, toroidal orbifolds are singular limits of certain Calabi-Yau manifolds.
Conversely, one can realize smooth manifolds like Calabi-Yau manifolds by blowing up orbifold singularities.
However, it is very difficult to study phenomenological aspects such as matter wave functions, explicit forms of Yukawa couplings and higher order 
couplings at a generic point in the moduli space of Calabi-Yau manifolds.\footnote{See, Refs. \cite{Ishiguro:2021drk,Ishiguro:2021ccl}, for the recent attempt in the context of heterotic string theory with standard embedding, where matter couplings can be extracted from moduli couplings.}
On the other hand, such analysis would be possible at nearby orbifold limits of Calabi-Yau manifolds.
Thus, it is important to study matter wave functions and Yukawa couplings on blow-up manifolds of the toroidal orbifolds. 
It would capture the quantitative aspects of Calabi-Yau compactifications.

For blow-up manifolds of $\mathbb{C}^N/Z_N$ with $N \geq 2$, 
metric and gauge fluxes have been obtained in Refs.~\cite{GrootNibbelink:2007lua,Leung:2019oln}, 
where the orbifold singular points are replaced by the Eguchi-Hanson spaces~\cite{Eguchi:1978xp}.
However, wave functions and their couplings on the blow-up manifolds have not been obtained yet.
On the other hand, blow-up of $T^2/Z_N$ orbifolds with bulk magnetic fluxes was studied in Ref.~\cite{Kobayashi:2019fma}, 
where the orbifold singular points are replaced by the part of $S^2$.
Then, matter wave functions, their couplings, and the flavor structure were studied \cite{Kobayashi:2019fma,Kobayashi:2019gyl}.

In addition to bulk gauge fluxes, localized fluxes are possible at singular points of orbifolds ~\cite{Lee:2003mc}\footnote{
Even if localized fluxes vanish at the tree level, they may be induced by loop effects~\cite{Lee:2003mc,Abe:2020vmv}. }.
They affect the number of zero-modes through the index theorem~\cite{Kobayashi} and their wave functions and couplings.
That is, localized gauge fluxes would play an important role in 4D low-energy effective field theory.
The purpose of this paper is to study the $T^2/Z_N$ orbifolds and these blow-up manifolds with 
both bulk and localized gauge fluxes.
We examine implications of localized gauge fluxes. 
It can be realized by a singular gauge transformation to remove $Z_N$ phases in the $Z_N$ twisted boundary condition.
Remarkably, such localized gauge fluxes induce new chiral zero modes as studied from the viewpoint of index theorem~\cite{Kobayashi}. 
In this followup paper, we examine the profile of these chiral zero-mode wave functions in more detail. 
By computing their wave functions, we find that they correspond to localized modes at the orbifold singular point of $T^2/Z_N$ orbifolds. 
We also study Yukawa couplings among bulk and localized zero-modes. 

This paper is organized as follows.
In Sec.~\ref{sec:construction}, we review the construction of blow-up manifolds of $T^2/Z_N$ orbifolds.
In Sec.~\ref{sec:bulkmode}, we compute wave functions of bulk zero modes on the blow-up manifold with magnetic fluxes in more detail, 
where only $Z_N$ invariant modes have been analyzed in Ref. \cite{Kobayashi:2019fma}. 
In addition, in Sec.~\ref{sec:localmode}, we study wave functions of new zero modes, which correspond to localized zero modes on $T^2/Z_N$ orbifolds.
We also discuss Yukawa couplings among  bulk and localized zero-modes  in Sec.~\ref{sec:Yukawa}.
We conclude this study in Sec.~\ref{sec:conclusion}.
In Appendix~\ref{ap:normalbulk}, we show the detailed calculation of normalization of bulk zero modes.
In Appendix~\ref{ap:normallocal}, we show the detailed calculation of normalization of localized zero modes.

% ------------------------------------------------------ %
% ------------------------------------------------------ %
% ------------------------------------------------------ %
% ------------------------------------------------------ %

\section{Blow-up manifold of $T^2/Z_N$ orbifold}
\label{sec:construction}

In this section, we briefly review the construction of blow-up manifolds of $T^2/Z_N$ orbifolds~\cite{Kobayashi:2019fma}.

Firstly, a two-dimensional (2D) torus $T^2$ can be constructed as division of a complex plane $\mathbb{C}$ by a 2D lattice $\Lambda$, 
i.e., $\mathbb{C}/\Lambda$. 
The 2D lattice itself is generated by two lattice vectors.
Here, one lattice vector is normalized as $1$ on the complex plane, 
and the other becomes a complex number $\tau$ so-called the complex structure modulus of $T^2$.
We define a complex coordinate of $T^2$, $z$, such that the identifications are $z \sim z+1 \sim z+\tau$.
The torus $T^2$ is flat, and its curvature vanishes.
The area of $T^2$ becomes ${\rm Im}\tau$.

Second, $T^2/Z_N$ orbifolds can be constructed by further identifying a $Z_N$ twisted point $\rho z$ ($\rho=e^{2\pi i/N}$) with $z$,~i.e., $\rho z \sim z$, where $N$ must be either $2$, $3$, $4$, or $6$ such that a lattice point transforms another lattice point under the $Z_N$ twist.
Note that $\tau$ is constrained as $\tau=\rho$ for $N=3,4,6$, while it is not constrained for $N=2$.
The $T^2/Z_N$ orbifolds have fixed points $z_I $ for $Z_N$ twist up to the $T^2$ translation,~i.e., 
\begin{align}
\rho z_I + u + v\tau = z_I \qquad
(\exists\, u, v \in \mathbb{Z}).
\end{align}
They correspond to orbifold singular points with the curvature determined by $\theta/ 2\pi$, where 
$\theta$ denotes the deficit angle around the singular point.

\begin{figure}[H]
    \centering
    \begin{minipage}{7cm}
    \centering
    \includegraphics[bb=0 0 750 700,width=5cm]{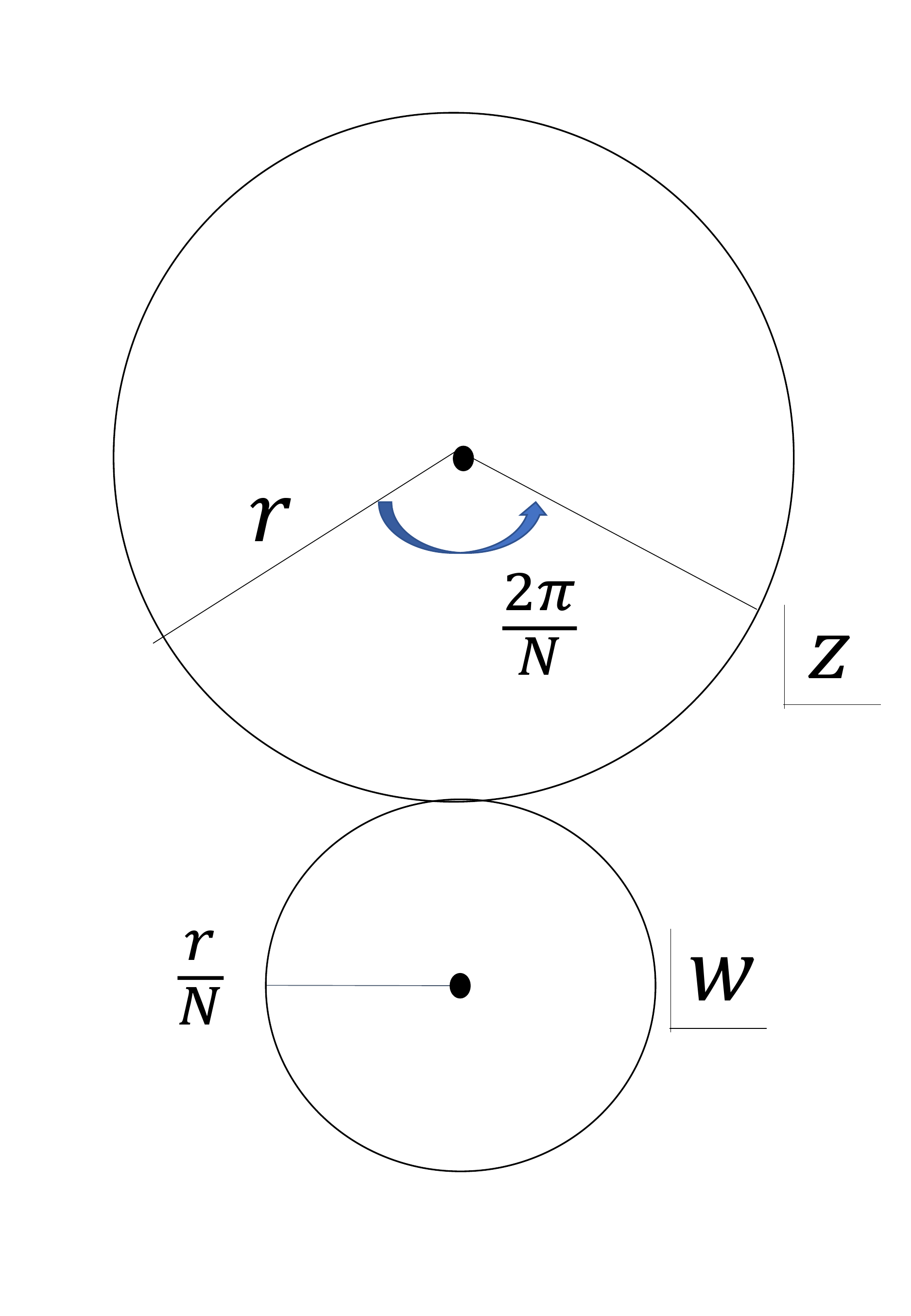}
    \end{minipage}
    \begin{minipage}{7cm}
    \centering
    \includegraphics[bb=0 0 500 375,width=5cm]{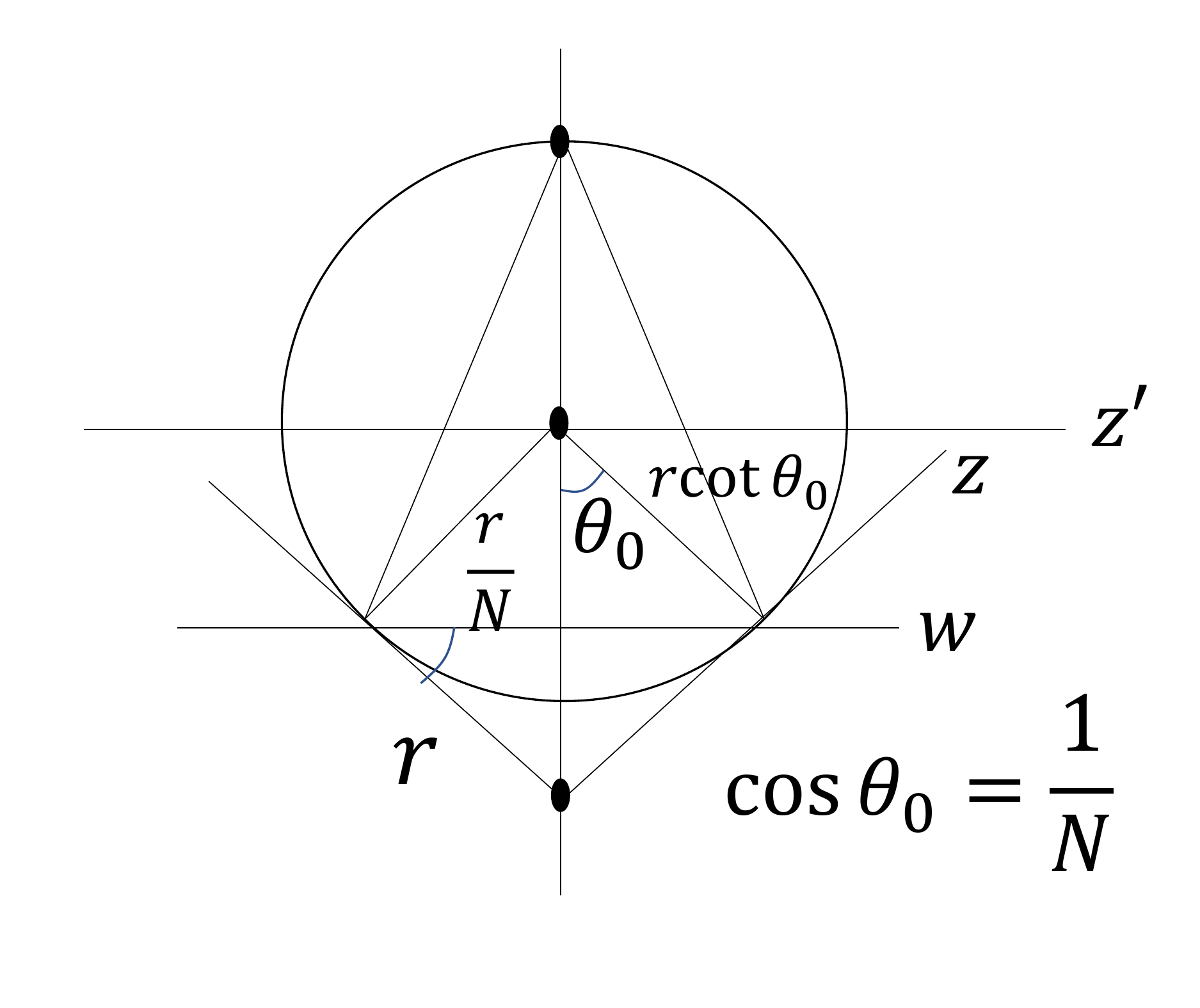}
    \end{minipage}
    \caption{The left figure shows the development of the cone around a singular point of $T^2/Z_N$ orbifold. The right figure shows the cross section of the cone and the $S^2$ with radius $R =r/\sqrt{N^2-1}$. Here, $z$ and $z'$ denote the coordinates of $T^2/Z_N$ and $S^2$, respectively, and they are related through the coordinate $w$.}
    \label{fig:ZN}
\end{figure}

\begin{figure}[H]
    \centering
    \includegraphics[bb=0 0 500 600,width=5cm]{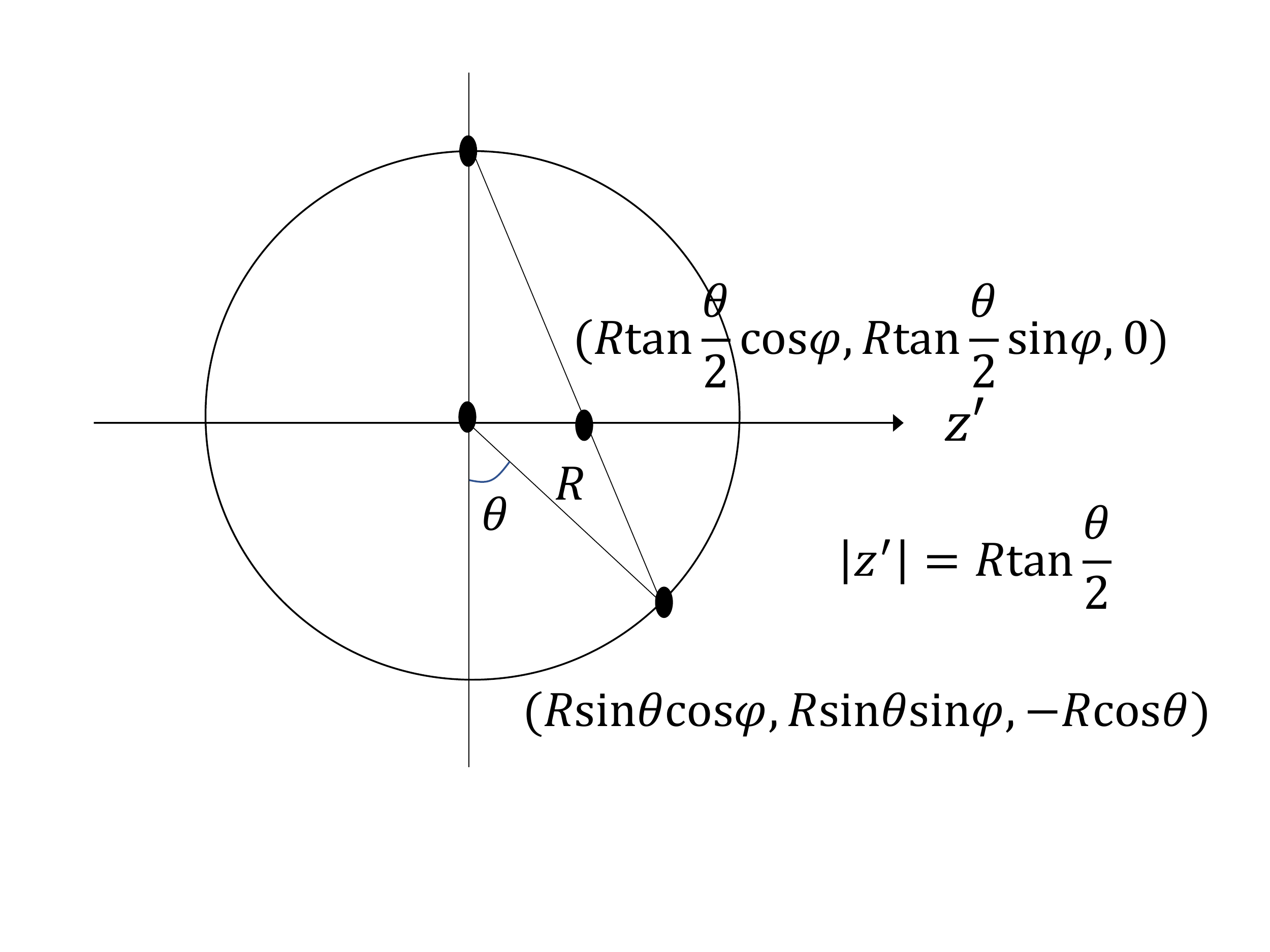}
    \caption{The coordinate of $\mathbb{CP}^1 \simeq S^2$.}
    \label{fig:coS2}
\end{figure}

Finally, blow-up manifolds of $T^2/Z_N$ orbifolds are constructed by replacing the cone around the orbifold singular point with the part of sphere $S^2$~\cite{Kobayashi:2019fma}, as shown in Figure~\ref{fig:ZN}.
Figure~\ref{fig:ZN} shows the case that the deficit angle around the singular point is $2\pi(N-1)/N$, and we replace the cone whose slant height is $r$ with $(N-1)/2N$-part of $S^2$ whose radius is $R =r/\sqrt{N^2-1}$.
Note that the curvature of $S^2$ is $\chi(S^2)=2$, and then this replacement does not change the topological invariant number.
The left figure shows the development of the cone, and the right figure shows the cross section of the cone and $S^2$ with the radius $R =r/\sqrt{N^2-1}$.
Here, $z'$ denotes the complex coordinate of $\mathbb{CP}^1 \simeq S^2$, defined by projecting a point of $S^2$ into the complex plane passing through the center of $S^2$ from the north pole of $S^2$ as shown in Figure \ref{fig:coS2}, while $z$ denotes the complex coordinate of $T^2/Z_N$ orbifold.
Note that the definition of $z'$ is different from that in Ref.~\cite{Kobayashi:2019fma}.
They are related through the coordinate $w$,~i.e., $z|_{z=re^{i\varphi/N}} \leftrightarrow w= \frac{N+1}{N}z'|_{z'=\frac{r}{N+1}e^{i\varphi}}$.

In the following sections, we show chiral zero-mode wave functions on magnetized blow-up manifolds, which can be obtained by smoothly connecting wave functions on the magnetized $T^2/Z_N$ orbifold with those on the magnetized $S^2$.

% ------------------------------------------------------ %
% ------------------------------------------------------ %
% ------------------------------------------------------ %
% ------------------------------------------------------ %

\section{Bulk zero-mode wave functions}
\label{sec:bulkmode}

% ------------------------------------------------------ %
% ------------------------------------------------------ %

\subsection{Magnetized $T^2/Z_N$ orbifold}
\label{subsec:wavT2ZN}

In this subsection, we review chiral zero-mode wave functions on the magnetized $T^2/Z_N$ orbifolds.

Firstly, we analyze magnetized $T^2$ with $U(1)$ magnetic flux.
The magnetic flux must be quantized by the Dirac quantization condition,
\begin{align}
\int_{T^2} \frac{F}{2\pi} = M \in \mathbb{Z}.
\end{align}
This magnetic flux is given by the field strength,
\begin{align}
    \frac{F}{2\pi} = \frac{i}{2} \frac{M}{{\rm Im}\tau} dz \land d\bar{z},
\end{align}
and it is induced by the vector potential,
\begin{align}
    A = -\frac{i}{2} \frac{\pi M}{{\rm Im}\tau}\bar{z} dz + \frac{i}{2} \frac{\pi M}{{\rm Im}\tau}z d\bar{z}.
\end{align}
A 2D spinor $\psi_{T^2}^{(\alpha_1,\alpha_{\tau})}=(\psi_{T^2,+}^{(\alpha_1,\alpha_{\tau})}, \psi_{T^2,-}^{(\alpha_1,\alpha_{\tau})})^T$ on the magnetized $T^2$ with $U(1)$ charge, $q=1$, satisfies the following boundary conditions:
\begin{align}
    \begin{array}{ll}
        \psi_{T^2,\pm}^{(\alpha_1,\alpha_{\tau}),M}(z+1) = e^{2\pi i \alpha_1} e^{i \chi_1(z)} \psi_{T^2,\pm}^{(\alpha_1,\alpha_{\tau}),M}(z), & \chi_1(z) = \pi M\frac{{\rm Im}z}{{\rm Im}\tau}, \\
        \psi_{T^2,\pm}^{(\alpha_1,\alpha_{\tau}),M}(z+\tau) = e^{2\pi i \alpha_{\tau}} e^{i \chi_{\tau}(z)} \psi_{T^2,\pm}^{(\alpha_1,\alpha_{\tau}),M}(z), & \chi_{\tau}(z) = \pi M\frac{{\rm Im}(\bar{\tau}z)}{{\rm Im}\tau}.
    \end{array}
    \label{eq:BCT2}
\end{align}
Here, $(\alpha_1, \alpha_{\tau})$ denote the degree of freedom of Scherk-Schwarz (SS) phases.
We do not consider Wilson line (WL) phases since they can be rewritten by SS phases~\cite{Abe:2013bca}.
The zero-mode equation for the Dirac operator $\slashed{D}$ on the magnetized $T^2$,
\begin{align}
 i\slashed{D} \psi_{T^2}^{(\alpha_1,\alpha_{\tau}),M}(z) = 0, 
\end{align}
can be written by 
gamma matrices, $\gamma_z$ and $\gamma_{\bar{z}}$, and covariant derivatives, $D_z$ and $D_{\bar{z}}$, 
\begin{align}
    \begin{array}{l}
         \gamma_z =
         \begin{pmatrix}
         0 & 2 \\ 0 & 0
         \end{pmatrix}, \quad
         \gamma_{\bar{z}} =
         \begin{pmatrix}
         0 & 0 \\ 2 & 0
         \end{pmatrix}, \\
         D_z = \partial_z -iA_z, \quad D_{\bar{z}} = \partial_{\bar{z}} -iA_{\bar{z}}.
    \end{array}
\end{align}
When $M$ is positive, only $\psi_{T^2,+}^{(\alpha_1,\alpha_{\tau}),M}$ has zero-mode solutions and there are $M$ number of zero-modes,
\begin{align}
    \begin{array}{ll}
        \psi_{T^2,+,0}^{(\alpha_1,\alpha_{\tau}),M}(z) = \sum_{j=0}^{M-1} \psi_{T^2,+,0}^{(j+\alpha_1,\alpha_{\tau}),M}(z), \\
        \psi_{T^2,+,0}^{(j+\alpha_1,\alpha_{\tau}),M}(z) = e^{-\frac{\pi M}{2{\rm Im}\tau}|z|^2} g^{(j+\alpha_1,\alpha_{\tau}),M}_{T^2}(z),  \\
        g^{(j+\alpha_1,\alpha_{\tau}),M}_{T^2}(z) = {\cal N}_{T^2} e^{\frac{\pi M}{2{\rm Im}\tau}z^2} e^{2\pi i\frac{j+\alpha_1}{M}\alpha_{\tau}} \vartheta
        \begin{bmatrix}
        \frac{j+\alpha_1}{M} \\ -\alpha_{\tau}
        \end{bmatrix}
        (Mz,M\tau), \quad j \in \mathbb{Z}/M\mathbb{Z},
    \end{array} \label{eq:wavT2}
\end{align}
where $g^{(j+\alpha_1,\alpha_{\tau}),M}_{T^2}(z)$ denotes the holomorphic function of $z$ including a normalization factor, ${\cal N}_{T^2}$, determined by the normalization condition, 
\begin{align}
\int_{T^2} dzd\bar{z} (\psi_{T^2,+,0}^{(j+\alpha_1,\alpha_{\tau}),M})^{\ast} \psi_{T^2,+,0}^{(k+\alpha_1,\alpha_{\tau}),M} = \delta_{j,k}.
\end{align}
We introduced the Jacobi theta function defined by
\begin{align}
    \vartheta
    \begin{bmatrix}
     a \\ b
    \end{bmatrix}
    (\nu,\tau) =
    \sum_{\ell \in \mathbb{Z}}
    e^{\pi i (a+\ell)^2\tau}
    e^{2\pi i (a+\ell)(\nu+b)}.
\end{align}
When $M$ is negative, on the other hand, only $\psi_{T^2,-}^{(\alpha_1,\alpha_{\tau}),M}$ has $|M|$ number of zero-mode solutions and they are given by replacing $(M,z,\tau) \rightarrow (|M|, \bar{z}, \bar{\tau})$ in Eq.~(\ref{eq:wavT2}).
Hereafter, we consider $M>0$ case, that is, zero-modes have positive chirality.

Next, let us see chiral zero-mode wave functions on the magnetized $T^2/Z_N$ orbifolds~\cite{Abe:2008fi,Abe:2013bca}.
In addition to Eq.~(\ref{eq:BCT2}), they further satisfy the following boundary condition:
\begin{align}
    \begin{array}{l}
         \psi_{T^2/Z_N^m,+}^{(\alpha_1,\alpha_{\tau}),M}(\rho z) = \rho^m \psi_{T^2/Z_N^m,+}^{(\alpha_1,\alpha_{\tau}),M}(z), \\
         \psi_{T^2/Z_N^m,-}^{(\alpha_1,\alpha_{\tau}),M}(\rho z) = \rho^{m+1} \psi_{T^2/Z_N^m,-}^{(\alpha_1,\alpha),M}(z),
    \end{array}
    \label{eq:BCT2ZN}
\end{align}
where $\rho \equiv e^{2\pi i/N}$ and $m \in \mathbb{Z}/N\mathbb{Z}$ denotes the $Z_N$ charge.
Then, wave functions on the magnetized $T^2/Z_N$ orbifold can be expanded by wave functions on the magnetized $T^2$ as
\begin{align}
    \begin{array}{l}
         \psi_{T^2/Z_N^m,+}^{(\alpha_1,\alpha_{\tau}),M}(z) = {\cal N}_{T^2/Z_N} \sum_{k=0}^{N-1} \rho^{-km} \psi_{T^2,+}^{(\alpha_1,\alpha_{\tau}),M}(\rho^k z), \\
         \psi_{T^2/Z_N^m,-}^{(\alpha_1,\alpha_{}\tau),M}(z) = {\cal N}_{T^2/Z_N} \sum_{k=0}^{N-1} \rho^{-k(m+1)} \psi_{T^2,-}^{(\alpha_1,\alpha_{\tau}),M}(\rho^k z),
    \end{array}
\end{align}
where ${\cal N}_{T^2/Z_N}$ similarly denotes the normalization factor.
In particular, chiral zero-mode wave functions on the magnetized $T^2/Z_N$ orbifold can be expressed as
\begin{align}
    \begin{array}{l}
         \psi_{T^2/Z_N^m,+,0}^{(\alpha_1,\alpha_{\tau}),M}(z) = \sum_{j} \psi_{T^2/Z_N^m,+,0}^{(j+\alpha_1,\alpha_{\tau}),M}(z), \\
         \psi_{T^2/Z_N^m,+,0}^{(j+\alpha_1,\alpha_{\tau}),M}(z) = e^{-\frac{\pi M}{2{\rm Im}\tau}|z|^2} h^{(j+\alpha_1,\alpha_{\tau}),M}_{T^2/Z_N^m}(z), \\
         h^{(j+\alpha_1,\alpha_{\tau}),M}_{T^2/Z_N^m}(z) = {\cal N}_{T^2/Z_N} \sum_{k=0}^{N-1} \rho^{-km} g^{(j+\alpha_1,\alpha_{\tau}),M}_{T^2}(\rho^k z),
    \end{array}
    \label{eq:wavT2ZN}
\end{align}
where $h^{(j+\alpha_1,\alpha_{\tau}),M}_{T^2/Z_N^m}(z)$ also denotes the holomorphic function of $z$.
The number of zero modes can be determined by the magnetic flux $M$, and the boundary conditions $(\alpha_1, \alpha_{\tau};m)$.

% ------------------------------------------------------ %
% ------------------------------------------------------ %

\subsection{Magnetized $S^2$}
\label{subsec:wavS2}

In this subsection, we review chiral zero-mode wave functions on the magnetized $S^2$ with $U(1)$ magnetic flux~\cite{Conlon:2008qi}.
The magnetic flux is quantized as
\begin{align}
\int_{S^2} \frac{F'}{2\pi} = M' \in \mathbb{Z}.
\end{align}
This magnetic flux is given by the field strength,
\begin{align}
    \frac{F'}{2\pi} = \frac{i}{2\pi} \frac{R^2 M'}{(R^2 + |z'|^2)^2} dz' \land d\bar{z}',
\end{align}
and it is induced by the vector potential,
\begin{align}
    A' = -\frac{i}{2} \frac{M'}{R^2 + |z'|^2}\bar{z}' dz' +  \frac{i}{2} \frac{M'}{R^2 + |z'|^2}z' d\bar{z}'.
\end{align}
The zero-mode equation of a 2D spinor $\psi_{S^2}=(\psi_{S^2,+}, \psi_{S^2,-})^T$ for the Dirac operator $\slashed{D}'$ on the magnetized $S^2$,
\begin{align}
 i\slashed{D}' \psi_{S^2}(z') = 0,
\end{align}
can be written by gamma matrices, ${\gamma'}_{z'}$ and ${\gamma'}_{\bar{z}'}$, and covariant derivatives, ${D'}_{z'}$ and ${D'}_{\bar{z}'}$, 
\begin{align}
    \begin{array}{l}
         {\gamma'}_{z'} =
         \begin{pmatrix}
         0 & \frac{R^2 + |z'|^2}{R} \\ 0 & 0
         \end{pmatrix}, \quad
         {\gamma'}_{\bar{z}'} =
         \begin{pmatrix}
         0 & 0 \\ \frac{R^2 + |z'|^2}{R} & 0
         \end{pmatrix}, \\
         {D'}_{z'} = \partial_{z'} + \frac{i}{2}\sigma_3 {\omega'}_{z'} -i{A'}_{z'} , \quad {D'}_{\bar{z}'} = \partial_{\bar{z}'} + \frac{i}{2}\sigma_3 {\omega'}_{\bar{z}'} -i{A'}_{\bar{z}'},
    \end{array}
\end{align}
where
\begin{align}
    \omega' = -\frac{i}{2} \frac{2}{R^2 + |z'|^2}\bar{z}' dz' + \frac{i}{2} \frac{2}{R^2 + |z'|^2}z' d\bar{z}'
\end{align}
denotes the spin connection, indicating a presence of the curvature of $S^2$: 
\begin{align}
\int_{S^2} \frac{R'}{2\pi} = \chi(S^2) = 2.
\end{align}
Note that the spin connection is the same functional form as the gauge potential, where the former and the latter are proportional to the curvature $2$ and the magnetic flux $M'$, respectively.
In particular, zero-modes of $\psi_{S^2,+}$ can be expressed as
\begin{align}
    \psi_{S^2,+,0}^{M'}(z') = \frac{f_{S^2}^{M'}(z')}{(R^2 + |z'|^2)^{\frac{M'-1}{2}}},
    \label{eq:wavS2}
\end{align}
where $f_{S^2}^{M'}(z')$ denotes the holomorphic function of $z'$.
They exist if $M'$ is positive and $f_{S^2}^{M'}(z')$ can be written by $(M'-1)$-polynomials due to the normalization condition of them on the magnetized $S^2$.

% ------------------------------------------------------ %
% ------------------------------------------------------ %

\subsection{Singular gauge transformation}
\label{subsec:reT2ZN}

Since we have obtained both wave functions on the magnetized $T^2/Z_N$ orbifold and $S^2$, the next step is to connect them smoothly.
However, when wave functions in Eq.~(\ref{eq:wavT2ZN}) go along the circle of the base of the cone around the singular point $z_I=0$, $Z_N$ phase $\rho^m$ appears from them due to Eq.~(\ref{eq:BCT2ZN}), while no phase appears from wave functions in Eq.~(\ref{eq:wavS2}) when they go along the same circle on the sphere.
Thus, in this subsection, we consider removing the $Z_N$ phase from Eq.~(\ref{eq:BCT2ZN}) by a (singular) gauge transformation, as in the case that SS phases and WL phase are related through the gauge transformation.

In Ref.~\cite{Dolan:2020sjq}, wave functions on magnetized $S^2$ with vortices have been studied.
In such a theory, the singular gauge transformation at $z'=0$ is generated by $U = {z'}/{\bar{z}'}$ and the field strength is only modified at $z'=0$.
If we apply this to orbifold singular points, we can modify the $Z_N$ boundary conditions in Eq.~(\ref{eq:BCT2ZN}).
Instead, it may induce localized fluxes.
Furthermore, the case that localized fluxes are introduced at orbifold fixed points has been studied in Ref.~\cite{Lee:2003mc}, and it is related to
\begin{align}
    \begin{array}{l}
        \psi^{(\frac{1}{2},\frac{1}{2}),1}_{T^2/Z_N^1}(z) =  \psi^{(\frac{1}{2},\frac{1}{2}),1}_{T^2}(z) =  e^{-\frac{\pi}{2{\rm Im}\tau}|z|^2} g_1(z) \\
        g_1(z) \equiv g^{(\frac{1}{2},\frac{1}{2}),1}_{T^2} = e^{\frac{\pi}{2{\rm Im}\tau}z^2} e^{\frac{\pi i}{2}} \vartheta
        \begin{bmatrix}
        \frac{1}{2} \\ -\frac{1}{2}
        \end{bmatrix}
        (z,\tau)
    \end{array}.
\end{align}
(See also Ref.~\cite{Polchinski,Sakamoto:2020pev}.)
Then, we define the singular gauge transformation as
\begin{align}
    \begin{array}{l}
        A \rightarrow \tilde{A} = A + \delta A, \\
        \delta A = i U_{\xi^F_0} d U_{\xi^F_0}^{-1} 
        = -i \frac{\xi^F_0}{2} \frac{g^{(1)}_1(z)}{g_1(z)} dz + i \frac{\xi^F_0}{2} \frac{\overline{g^{(1)}_1(z)}}{\overline{g_1(z)}} d\bar{z}
        \simeq -i \frac{\xi^F_0}{2} \frac{1}{z} dz + i \frac{\xi^F_0}{2} \frac{1}{\bar{z}} d\bar{z},
        \label{eq:Asingauge}
    \end{array}
\end{align}
with
\begin{align}
    U_{\xi^F_0} = \left( \frac{\psi^{(\frac{1}{2},\frac{1}{2}),1}_{T^2/Z_N^1}(z)}{\overline{\psi^{(\frac{1}{2},\frac{1}{2}),1}_{T^2/Z_N^1}(z)}} \right)^{\frac{\xi^F_0}{2}}= \left( \frac{g_1(z)}{\overline{g_1(z)}} \right)^{\frac{\xi^F_0}{2}}
        \simeq \left( \frac{g^{(1)}_1(0)z}{\overline{g^{(1)}_1(0)z}} \right)^{\frac{\xi^F_0}{2}}, \label{eq:singunitary}
\end{align}
where the rightest side in Eq. (\ref{eq:Asingauge}) shows the approximation around $z_I=0$ and $g^{(n)}_1(z) \equiv \frac{d^n g_1(z)}{dz^n}$.
Because of the above singular gauge transformation, the field strength is modified as
\begin{align}
    \begin{array}{l}
        \frac{F}{2\pi} \rightarrow \frac{\tilde{F}}{2\pi} = \frac{F}{2\pi} + \frac{\delta F}{2\pi}, \\
        \frac{\delta F}{2\pi} = i \xi^F_0 \delta(z) \delta(\bar{z}) dz \land d\bar{z},
    \end{array}
\end{align}
and it induces the localized flux $\xi^F_0/N$ at $z_I=0$.
The detailed analysis of the localized flux is discussed in Ref.~\cite{Kobayashi}.
Similarly, since the curvature at $z_I=0$ becomes $\xi^R_0/N = (N-1)/N$ (while the curvature at the other points except for fixed points remains $0$) and also the spin connection is the same form as the gauge potential by replacing the magnetic flux with the curvature in the case of $S^2$, we define the singular gauge transformation for the spin connection as well:
\begin{align}
    \begin{array}{l}
        w=0 \rightarrow \tilde{w} = w + \delta w = \delta w, \\
        \delta w = i U_{\xi^R_0} d U_{\xi^R_0}^{-1}, \qquad U_{\xi^R_0} = \left( \frac{\psi^{(\frac{1}{2},\frac{1}{2}),1}_{T^2/Z_N^1}(z)}{\overline{\psi^{(\frac{1}{2},\frac{1}{2}),1}_{T^2/Z_N^1}(z)}} \right)^{\frac{\xi^R_0}{2}}.
    \end{array}
\end{align}
According to the singular gauge transformation for gauge potential and spin connection, wave functions can be rewritten as
\begin{align}
    \begin{array}{l}
         \tilde{\psi}_{T^2/Z_N^m,+}^{(\alpha_1,\alpha_{\tau})}(z) = U_{\xi^F_0} U_{\xi^R_0}^{-1/2} \psi_{T^2/Z_N^m, +}^{(\alpha_1,\alpha_{\tau})}(z) = \left( \frac{\psi^{(\frac{1}{2},\frac{1}{2}),1}_{T^2/Z_N^1}(z)}{\overline{\psi^{(\frac{1}{2},\frac{1}{2}),1}_{T^2/Z_N^1}(z)}} \right)^{\frac{\xi^F_0}{2}-\frac{1}{2}\frac{\xi^R_0}{2}} \psi_{T^2/Z_N^m, +}^{(\alpha_1,\alpha_{\tau})}(z),  \\
         \tilde{\psi}_{T^2/Z_N^m,-}^{(\alpha_1,\alpha_{\tau})}(z) = U_{\xi^F_0} U_{\xi^R_0}^{1/2} \psi_{T^2/Z_N^m, -}^{(\alpha_1,\alpha_{\tau})}(z) =  \left( \frac{\psi^{(\frac{1}{2},\frac{1}{2}),1}_{T^2/Z_N^1}(z)}{\overline{\psi^{(\frac{1}{2},\frac{1}{2}),1}_{T^2/Z_N^1}(z)}} \right)^{\frac{\xi^F_0}{2}+\frac{1}{2}\frac{\xi^R_0}{2}} \psi_{T^2/Z_N^m, -}^{(\alpha_1,\alpha_{\tau})}(z), 
    \end{array}
\end{align}
and then the boundary conditions in Eq.~(\ref{eq:BCT2ZN}) can be modified as
\begin{align}
    \begin{array}{l}
         \tilde{\psi}_{T^2/Z_N^m,+}^{(\alpha_1,\alpha_{\tau})}(\rho z) = \rho^{\xi^F_0 - \frac{1}{2}\xi^R_0 + m} \tilde{\psi}_{T^2/Z_N^m,+}^{(\alpha_1,\alpha_{\tau})}(z), \\
         \tilde{\psi}_{T^2/Z_N^m,-}^{(\alpha_1,\alpha_{\tau})}(\rho z) = \rho^{\xi^F_0 + \frac{1}{2}\xi^R_0 + m+1} \tilde{\psi}_{T^2/Z_N^m,-}^{(\alpha_1,\alpha)}(z). 
    \end{array}
    \label{eq:BCmodT2ZN}
\end{align}
Thus, when we consider $\xi^F_0 = \frac{N-1}{2} - m + \ell_0 N$, $\ell_0 \in \mathbb{Z}$, we can remove $Z_N$ phase from Eq.~(\ref{eq:BCmodT2ZN}), where we use $\xi^R_0 = N-1$.
In the next section, we discuss the physical meaning of the degree of freedom of localized flux $\ell_0$.
Note that the boundary conditions in Eq.~(\ref{eq:BCT2}) are also modified as
\begin{align}
    \begin{array}{ll}
        \tilde{\psi}_{T^2/Z_N^m,\pm}^{(\alpha_1,\alpha_{\tau})}(z+1) = e^{2\pi i (\alpha_1 + \frac{\xi^F_0}{2}\mp\frac{1}{2}\frac{\xi^R_0}{2})} e^{i \tilde{\chi}_1(z)} \tilde{\psi}_{T^2/Z_N^m,\pm}^{(\alpha_1,\alpha_{\tau})}(z), & \tilde{\chi}_1(z) = \pi (M+\xi^F_0\mp\frac{\xi^R_0}{2}) \frac{{\rm Im}z}{{\rm Im}\tau}, \\
        \tilde{\psi}_{T^2/Z_N^m,\pm}^{(\alpha_1,\alpha_{\tau})}(z+\tau) =e^{2\pi i (\alpha_{\tau} + \frac{\xi^F_0}{2}\mp\frac{1}{2}\frac{\xi^R_0}{2})} e^{i \tilde{\chi}_{\tau}(z)} \tilde{\psi}_{T^2/Z_N^m,\pm}^{(\alpha_1,\alpha_{\tau})}(z), & \tilde{\chi}_{\tau}(z) = \pi (M+\xi^F_0\mp\frac{\xi^R_0}{2}) \frac{{\rm Im}(\bar{\tau}z)}{{\rm Im}\tau}.
    \end{array}
\end{align}
In particular, chiral zero-mode wave functions in Eq.~(\ref{eq:wavT2ZN}) can be modified as
\begin{align}
    \begin{array}{rl}
        \tilde{\psi}_{T^2/Z_N^m,+,0}^{(j+\alpha_1,\alpha_{\tau}),M}(z) =& \left| g_1(z) \right|^{m-\ell_0N} e^{-\frac{\pi M}{2{\rm Im}\tau}|z|^2} \tilde{h}^{(j+\alpha_1,\alpha_{\tau}),M}_{T^2/Z_N^m}(z) \\
        \simeq& |z|^{m-\ell_0N} e^{-\frac{\pi M}{2{\rm Im}\tau}|z|^2} |g^{(1)}_1(0)|^{m-\ell_0N} \tilde{h}^{(j+\alpha_1,\alpha_{\tau}),M}_{T^2/Z_N^m}(z), \\
        \tilde{h}^{(j+\alpha_1,\alpha_{\tau}),M}_{T^2/Z_N^m}(z) =& {\cal N}_{T^2/Z_N}^j \left( g_1(z) \right)^{-m+\ell_0N} \sum_{k=0}^{N-1} \rho^{-km} g^{(j+\alpha_1,\alpha_{\tau}),M}_{T^2}(\rho^k z) \\
        \simeq& {\cal N}_{T^2/Z_N}^j N \frac{(g^{(j+\alpha_1,\alpha_{\tau}),M}_{T^2})^{(m)}(0)}{m!} (g^{(1)}_1(0))^{-m+\ell_0N} z^{\ell_0N},
    \end{array}
    \label{eq:wavmodT2ZN}
\end{align}
where we also show the approximation around $z_I=0$ at the lowest order.
Hereafter, we denote the coefficient shortly as
\begin{align}
    C^{j}_{0} \equiv {\cal N}_{T^2/Z_N}^j \frac{(g^{(j+\alpha_1,\alpha_{\tau}),M}_{T^2})^{(m)}(0)}{m!} \left( \frac{g^{(1)}_1(0)}{|g^{(1)}_1(0)|} \right)^{-m+\ell_0N}.
\end{align}

% ------------------------------------------------------ %
% ------------------------------------------------------ %

\subsection{Normalized wave functions of bulk zero modes}
\label{subsec:bulkmode}

Now, let us see chiral zero-mode wave functions on the magnetized blow-up manifold obtained by smoothly connecting ones on the magnetized $T^2/Z_N$ orbifold in Eq.~(\ref{eq:wavmodT2ZN}) with ones on the magnetized $S^2$ in Eq.~(\ref{eq:wavS2}) at the connecting points.
Note that the renewed point from Ref.~\cite{Kobayashi:2019fma} is using Eq.~(\ref{eq:wavmodT2ZN}) instead of Eq.~(\ref{eq:wavT2ZN}).
Then, we can treat wave functions with $Z_N$ charge $m$ more precisely.

The junction conditions are given by
\begin{align}
    \begin{array}{c}
        \tilde{\psi}_{T^2/Z^m_N,+,0}^{(j+\alpha_1,\alpha_{\tau}),M}(z) \biggl|_{z=r e^{i\varphi/N}} = \psi_{S^2,+,0}^{M'}(z') \biggl|_{z'=\frac{r}{N+1}e^{i\varphi}},  \\
        \frac{1}{e^{-i\frac{\varphi}{N}}} \frac{d \tilde{\psi}_{T^2/Z^m_N,+,0}^{(j+\alpha_1,\alpha_{\tau}),M}(z)}{d z} \biggl|_{z=r e^{i\varphi/N}} = \frac{1}{\frac{N+1}{N}e^{-i\varphi}} \frac{d \psi_{S^2,+,0}^{M'}(z')}{d z'}\biggl|_{z'=\frac{r}{N+1}e^{i\varphi}}, 
    \end{array}
    \label{eq:concon}
\end{align}
where the derivatives of their coordinates can be written as
\begin{align}
    \begin{array}{c}
        e^{-i\frac{\varphi}{N}} dz = e^{-i\frac{\varphi}{N}} \frac{\partial z}{\partial |z|} d|z| + e^{-i\frac{\varphi}{N}} \frac{\partial z}{\partial (\frac{\varphi}{N})} d(\frac{\varphi}{N}) = d|z| + i r d(\frac{\varphi}{N}), \\
        \frac{N+1}{N} e^{-i\varphi} dz' = \frac{N+1}{N} e^{-i\varphi} \frac{\partial z'}{\partial |z'|} d|z'| + \frac{N+1}{N} e^{-i\varphi} \frac{\partial z'}{\partial \varphi} d\varphi = \frac{N+1}{N} d|z'| + i \frac{r}{N} d\varphi.
    \end{array}
\end{align}
Indeed, we find that the following relations:
\begin{align}
    \begin{array}{c}
        \frac{N+1}{N} d|z'| = \frac{N+1}{N} \frac{\partial |z'|}{\partial \theta}d\theta = \frac{N+1}{N} \frac{R}{2 \cos^2 \frac{\theta_0}{2}} d\theta =  \frac{N+1}{N} \frac{R}{1+\cos \theta_0} d\theta = Rd\theta = d|z|, \\
        rd(\frac{\varphi}{N}) = \frac{r}{N} d\varphi,
    \end{array}
\end{align}
are satisfied at the connecting points, as seen in Figure~\ref{fig:ZN}.
From non-holomorphic parts of wave functions in Eqs.~(\ref{eq:wavmodT2ZN}) and (\ref{eq:wavS2}), the junction conditions in Eq.~(\ref{eq:concon}) 
provide
\begin{align}
    \frac{\pi r^2}{N {\rm Im}\tau} M + \frac{N-1}{2N} - \frac{m}{N} + \ell_0 = \frac{N-1}{2N}M',
    \label{eq:fluxcond}
\end{align}
and from holomorphic parts, the holomorphic function on the part of $S^2$ region $f_{S^2}^{M'}(z')$ can be determined as
\begin{align}
    f_{S^2}^{M'}(z') = {C'}^{j}_{0} z'^{\ell_0}, \quad {C'}^{j}_{0} = C^{j}_{0} N r^{m} e^{-\frac{\pi M}{2{\rm Im}\tau}r^2} \left( \frac{r}{N+1} \right)^{M'-1-\ell_0} \left( \frac{N-1}{2N} \right)^{-\frac{M'-1}{2}}.
\end{align}
Note that the holomorphicity of bulk modes with positive flux $M$ requires $\ell_0 \geq 0$; otherwise 
they will diverge at $z'=0$. 
The divergence induced by the negative localized flux $\ell_0$ would be removed by introducing vortices analyzed in Ref.~\cite{Dolan:2020sjq}, which is beyond the scope of this paper. 
In the following analysis, we focus on the $\ell_0 \geq 0$ case.
Therefore, chiral zero-mode wave functions on magnetized blow-up manifolds can be written as
\begin{align}
    \psi^{j}_{{\rm blow-up},0} = \left\{
    \begin{array}{ll}
        \frac{{C'}^{j}_{0} z'^{\ell_0}}{(R^2 + |z'|^2)^{\frac{M'-1}{2}}} & (|z'| \leq \frac{r}{N+1}) \\
        \left| g_1(z) \right|^{m-\ell_0N} e^{-\frac{\pi M}{2{\rm Im}\tau}|z|^2} \tilde{h}^{(j+\alpha_1,\alpha_{\tau}),M}_{T^2/Z_N^m}(z) & (r \leq |z|) \\
        \ \simeq C^{j}_{0} N |z|^{m-\ell_0N} e^{-\frac{\pi M}{2{\rm Im}\tau}|z|^2} z^{\ell_0N}
    \end{array}
    \right.. \label{eq:wavbulk}
\end{align}
The flux condition in Eq.~(\ref{eq:fluxcond}) is generalized from that in Ref.~\cite{Kobayashi:2019fma}; the left-hand side shows the cut out flux from $T^2/Z_N$ orbifold, which is the flux including the localized flux on the cone of $T^2/Z_N$ orbifold, while the right-hand side shows the embedded flux, which is the flux on the part of $S^2$.
The detailed meaning is discussed in Ref.~\cite{Kobayashi}.
To determine the normalization, we first calculate the following inner product,
\begin{align}
    G_{ij}
    =& \int_{{\rm blow-up\ manifold}} dz d\bar{z} \sqrt{|{\rm det}(g)|} ( \psi^{i}_{{\rm blow-up},0} )^{\ast} \psi^{j}_{{\rm blow-up},0} \notag \\
    =&\, \delta_{i,j} - \int_{0}^{r} d|z| |z| \int_{0}^{\frac{2\pi}{N}} d\varphi (C^{i}_{0})^{\ast} C^{j}_{0} N^2 |z|^{2m} e^{-\frac{\pi M}{{\rm Im}\tau}|z|^2} \notag \\
    &+ \int_{0}^{\frac{r}{N+1}} d|z'| |z'| \int_{0}^{2\pi} d\varphi \frac{4R^4}{\left( R^2 + |z'|^2 \right)^2} \frac{({C'}^{i}_{0})^{\ast} {C'}^{j}_{0} |z'|^{2\ell_0}}{\left( R^2 + |z'|^2 \right)^{M'-1}} \notag \\
    \simeq&\, \delta_{i,j} +  (C^{i}_{0})^{\ast} C^{j}_{0} \pi (r^2)^{m+1} B_{0}\,, 
    \label{eq:normalbulk}
\end{align}
with 
\begin{align}
    B_{0} \simeq& \left( \frac{N-1}{2N}(M'-\ell_0) \right)^{-1} \frac{1-\sum_{p=0}^{\ell_0} \frac{\Gamma(M'+1)}{\Gamma(M'-p+1) \Gamma(p+1)}  \left( \frac{N+1}{2N} \right)^{M'-p} \left( \frac{N-1}{2N} \right)^{p}}{\frac{\Gamma(M'+1)}{\Gamma(M'-\ell_0+1) \Gamma(\ell_0+1)}  \left( \frac{N+1}{2N} \right)^{M'-\ell_0} \left( \frac{N-1}{2N} \right)^{\ell_0}} + \left( - \frac{m+1}{N} \right)^{-1}. \notag
\end{align}
We next perform the unitary transformation for flavor index $j$,
\begin{align}
    \begin{array}{l}
        \psi^{j'}_{{\rm blow-up},0} = U_{j'j} \psi^{j}_{{\rm blow-up},0} \\
        U = \prod_{J} \left( U^{J(J+1)} \right) {\rm diag}(e^{-i{\rm arg}(C^{j}_{0})}) \\
        U^{J(J+1)} =
        \begin{pmatrix}
        1 & \ & \ & \ & \ \\
        \ & \ddots & \ & \ & \ \\
        \ & \ &\begin{array}{cc}
        \cos \theta_{J(J+1)} & - \sin \theta_{J(J+1)} \\
        \sin \theta_{J(J+1)} & \cos \theta_{J(J+1)}
        \end{array}& \ & \ \\
        \ & \ & \ & \ddots & \ \\
        \ & \ & \ & \ & \ & 1
        \end{pmatrix}, \quad \tan^2 \theta_{J(J+1)} =
        \frac{\sum_{I=1}^{J} |C^{I}_{0}|^2}{|C^{J+1}_{0}|^2}.
    \end{array} \label{eq:unitary}
\end{align}
Then, the inner product $(G)_{i'j'}$ can be rewritten as
\begin{align}
    \begin{array}{l}
        G\simeq 
        \begin{pmatrix}
        1 & \ & \ \\
        \ & \ddots & \ \\
        \ & \ & 1 + \sum_{j} |C^{j}_{0}|^2 \pi (r^2)^{m+1} B_0
        \end{pmatrix}
        . \label{eq:reG}
    \end{array}
\end{align}
Thus, by redefining the normalization factor for the last mode $j'=j'_{\text{max}}$ as ${\cal N'}_{T^2/Z_N}^{j'_{\text{max}}} = {\cal N}_{T^2/Z_N}^{j'_{\text{max}}} (1 + O((r^2)^{m+1}))^{-1/2}$, all of the above modes can be expressed by orthonormal basis.
The detailed calculation of Eq.~(\ref{eq:normalbulk}) is shown in Appendix~\ref{ap:normalbulk}.

So far, we have focused wave functions around $z_I=0$.
Finally, we discuss wave functions around any orbifold singular points $z_I$.
First, we define $Z \equiv z - z_I$ and then $z=Z+z_I$.
From the point of view of the coordinate $Z$, $z_I$ will be regarded as the WL phase, and it corresponds to SS phase through the gauge transformation. 
Then, $Z_N$ twisted boundary condition for $Z$ can be obtained from that for $z$ in Eq.~(\ref{eq:BCT2ZN}), through $\rho z_I +u + v \tau = z_I$ and Eq.~(\ref{eq:BCT2}) in addition to the above, as
\begin{align}
         \psi_{T^2/Z_N^{V_I},+}^{(\beta_1, \beta_{\tau}),M}(\rho Z) = \rho^{V_I} \psi_{T^2/Z_N^{V_I},+}^{(\beta_1, \beta_{\tau}),M}(Z), \label{eq:BCT2ZNzI}
\end{align}
where $(\beta_1, \beta_{\tau})$ and $V_I$ are respectively given by
\begin{align}
    \begin{array}{l}
        (\beta_1, \beta_{\tau}) \equiv (\alpha_1+My_I,\alpha_{\tau}-Mx_I) \ ({\rm mod}\ 1) \quad (z_I = x_I + \tau y_I,\  \exists x_I, y_I \in \mathbb{R}), \\
        V_I \equiv [N(u\alpha_1 + v\alpha_{\tau} + (uv + u y_I - v x_I ) M/2) + m] \ ({\rm mod}\ N). 
    \end{array}
\end{align}
Therefore, the above analysis is valid by just the following replacement:
\begin{align}
    \begin{array}{c}
        \ z \rightarrow Z,  \\
        (\alpha_1,\alpha_{\tau}) \rightarrow (\beta_1, \beta_{\tau}), \\
        \ m \rightarrow V_I.
    \end{array}
    \label{eq:replace}
\end{align}
Since the above replacement includes $z_I=0$ case, one can obtain the matter wave functions around any singular points $z_I$.

% ------------------------------------------------------ %
% ------------------------------------------------------ %
% ------------------------------------------------------ %
% ------------------------------------------------------ %

\section{Localized zero-mode wave functions}
\label{sec:localmode}

Now, let us see the physical meaning of the degree of freedom of localized flux $\ell_0$.
As shown in Ref.~\cite{Kobayashi}, this degree of freedom means that there exists $\ell_0$ number of new chiral zero modes on the magnetized orbifold as well as the blow-up manifold.
In this section, we study wave functions of the new chiral zero-modes.

\begin{figure}[H]
    \centering
    \includegraphics[bb=0 0 400 320,width=6.5cm]{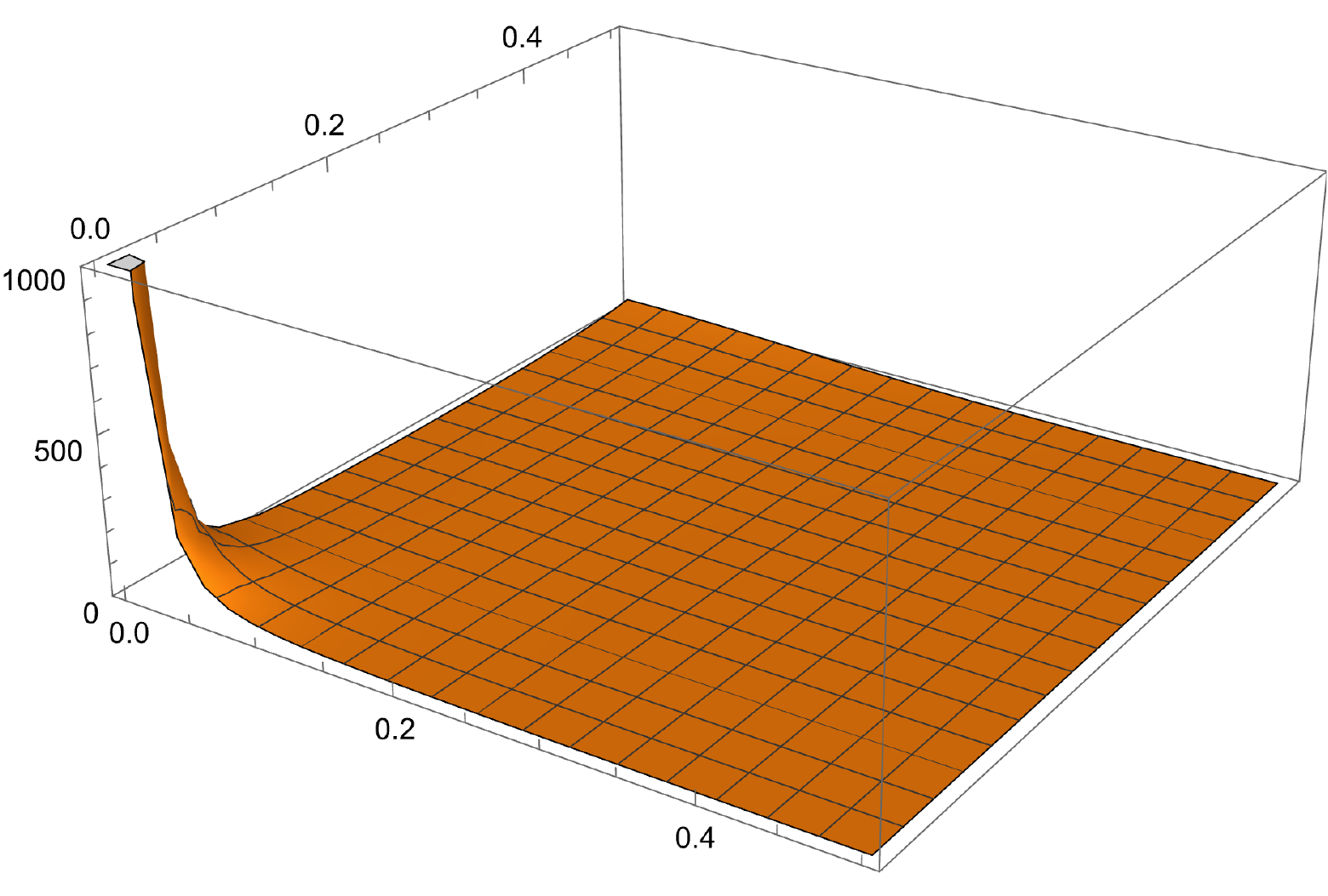}
    \caption{Probability density of unnormalized zero-mode wave function $|\tilde{\psi}^{a_0=\ell_0-1}_{T^2/Z_4,+,0}|^2$.}
    \label{fig:localizedmode}
\end{figure}

The bulk zero-mode wave functions, in the previous section, on the bulk region near the fixed point $z_I=0$ and the blow-up region are proportional to $z^{\ell_0N}$ and $z'^{\ell_0}$, respectively. 
It indicates that the new zero-mode wave functions on the bulk region near $z_I=0$ and the blow-up region will be proportional to $z^{a_0 N}$ and $z'^{a_0}$ for $a_0=0,...,\ell_0-1$, respectively.
Here, the factor $z^{\ell_0N}$ comes from the fact that the holomorphic function of the following wave function,
\begin{align}
    \psi^N_{T^2/Z_N^1}(z) \equiv \left( \psi^{(\frac{1}{2}, \frac{1}{2}),1}_{T^2/Z_N^1}(z) \right)^{N} = \left( \psi^{(\frac{1}{2}, \frac{1}{2}),1}_{T^2}(z) \right)^{N},
    \label{eq:psiN1}
\end{align}
is proportional to $z^N$ near the fixed point though it is $Z_N$ invariant, because it is made of the wave function with $Z_N$ charge $m=1$.
Note that its boundary condition is the same as that of wave functions with $M=N$, $(\alpha_1, \alpha_{\tau}) \equiv (\frac{N}{2}-[\frac{N}{2}], \frac{N}{2}-[\frac{N}{2}])$, and $m=0$,~i.e., $\psi^{(j+\frac{N}{2}-[\frac{N}{2}], \frac{N}{2}-[\frac{N}{2}]),N}_{T^2/Z_N^0}(z)$, and then the wave function in Eq.~(\ref{eq:psiN1}) can be expanded by these wave functions, where $[x]$ denotes the floor function. 
Thus, if the other wave function $\psi^N_{T^2/Z_N^0}(z)$, which has the same boundary condition of $\psi^{(j+\frac{N}{2}-[\frac{N}{2}], \frac{N}{2}-[\frac{N}{2}]),N}_{T^2/Z_N^0}(z)$, is constructed from $m=0$ mode,
we can obtain the new wave function whose holomorphic function is proportional to $z^{a_0 N}$ near $z_I=0$ by replacing $(\psi^N_{T^2/Z_N^1}(z))^{\ell_0-a_0}$ with $(\psi^N_{T^2/Z_N^0}(z))^{\ell_0-a_0}$.
Indeed, the zero-mode number of $\psi^{(j+\frac{N}{2}-[\frac{N}{2}], \frac{N}{2}-[\frac{N}{2}]),N}_{T^2/Z_N^0}(z)$ is just two, indicating that there exists the other zero-mode which is different from Eq.~(\ref{eq:psiN1}) which can be expanded by $\psi^{(j+\frac{N}{2}-[\frac{N}{2}], \frac{N}{2}-[\frac{N}{2}]),N}_{T^2/Z_N^0}(z)$.
Then, we can obtain $\psi^N_{T^2/Z_N^1}(z)$ as
\begin{align}
    \psi^N_{T^2/Z_N^0}(z)
    &\equiv e^{-\frac{\pi N}{2{\rm Im}\tau}|z|^2} h_0^N(z) \notag \\
    &\equiv \left\{
    \begin{array}{ll}
        \left( \psi^{(0,0),1}_{T^2/Z_N^0}(z) \right)^{N} = \left( \psi^{(0,0),1}_{T^2}(z) \right)^{N} & (N=2,4) \\
        \left( \psi^{(\frac{1}{6},\frac{1}{6}),1}_{T^2/Z_N^0}(z) \right)^{N} = \left( \psi^{(\frac{1}{6},\frac{1}{6}),1}_{T^2}(z) \right)^{N} & (N=3) \\
        \left( \psi^{(0,0),2}_{T^2/Z_N^0}(0) \psi^{(0,0),2}_{T^2/Z_N^0}(z) \right)^{N/2} & (N=6) %\\
    \end{array}
    \right.,
\end{align}
with
\begin{align}
    \psi^{(0,0),2}_{T^2/Z_N^0}(z) = \sqrt{\frac{\sqrt{3}+1}{2\sqrt{3}}} e^{-\pi i/8} \psi^{(0,0),2}_{T^2}(z) + \sqrt{\frac{\sqrt{3}-1}{2\sqrt{3}}} e^{\pi i/8} \psi^{(1,0),2}_{T^2}(z). \notag
\end{align}
Therefore, the  $\ell_0$ number of new chiral zero-mode wave functions can be expressed as
\begin{align}
    \begin{array}{rl}
        \tilde{\psi}^{a_0}_{T^2/Z_N,+,0} &\equiv {\cal N}^{a_0}_{T^2/Z_N} \left( \frac{\psi^N_{T^2/Z_N^0}(z)}{\psi^N_{T^2/Z_N^1}(z)} \right)^{\ell_0-a_0} \tilde{\psi}_{T^2/Z_N^m,+,0}^{(\alpha_1,\alpha_{\tau})}(z)  \\
         &\simeq C^{a_0}_{0} N |z|^{m-\ell_0N} e^{-\frac{\pi M}{2{\rm Im}\tau}|z|^2} z^{a_0 N}, 
    \end{array}
    \label{eq:wavloc}
\end{align}
where the coefficient $C^{a_0}$ is given by
\begin{align}
    C^{a_0}_{0} \equiv {\cal N}^{a_0}_{T^2/Z_N} \left( \frac{h_{0}^N(0)}{(g^{(1)}_{1}(0))^N} \right)^{\ell_0-a_0} \sum_{j} C^{j}_{0}.
\end{align}
Note that the non-holomorphic part of Eq.~(\ref{eq:wavloc}) does not change from that of Eq.~(\ref{eq:wavmodT2ZN}).
These new zero-modes diverge at the singular point $z_I=0$, while they are suppressed as they go away from the singular point, as shown in Figure~\ref{fig:localizedmode}.
That is, these new zero-modes correspond to localized modes around the singular point ($z_I=0$).
Although these localized modes diverge at $z_I=0$, they can be regularized by replacing the cone around $z_I=0$ with the part of $S^2$.
In other words, to calculate their normalization, we consider their wave functions on the magnetized blow-up manifold.
As in the previous section, through the connecting connection in Eq.~(\ref{eq:concon}), the wave functions on the magnetized blow-up manifold, which correspond to localized modes on the orbifold, can be written as
\begin{align}
    \psi^{a_0}_{{\rm blow-up},0} = \left\{
    \begin{array}{ll}
        \frac{{C'}^{a_0}_{0} z'^{a_0}}{(R^2 + |z'|^2)^{\frac{M'-1}{2}}} & (|z'| \leq \frac{r}{N+1}) \\
        \left| g_1(z) \right|^{m-\ell_0N} e^{-\frac{\pi M}{2{\rm Im}\tau}|z|^2}  {\cal N}^{a_0}_{T^2/Z_N} \left( \frac{h_0^N(z)}{(g_1(z))^N} \right)^{\ell_0-a_0} \sum_{j}\tilde{h}^j(z) & (r \leq |z|) \\
        \ \simeq C^{a_0}_{0} N |z|^{m-\ell_0N} e^{-\frac{\pi M}{2{\rm Im}\tau}|z|^2} z^{a_0 N} 
    \end{array}
    \right., \label{eq:wavlocalblowup}
\end{align}
where the coefficient ${C'}^{a_0}_{0}$ is given by
\begin{align}
    {C'}^{a_0}_{0} = C^{a_0}_{0} N r^{m-(\ell_0-a_0)N} e^{-\frac{\pi M}{2{\rm Im}\tau}r^2} \left( \frac{r}{N+1} \right)^{M'-1-a_0} \left( \frac{N-1}{2N} \right)^{-\frac{M'-1}{2}}.
\end{align}
Furthermore, since these wave functions are suppressed as they go away from the orbifold singular point, it has little effect on the result of inner product that we use an approximation form in all of the bulk region and also expand the integral region to $|z| \rightarrow \infty$.
Under this approximation, it turns out that the $\ell_0$ number of new zero modes are orthogonal to each other and also orthogonal to all of the bulk zero modes by using the following results:
\begin{align}
    \int_{0}^{\frac{2\pi}{N}} d{\rm arg}(z)\ z^{k N} = 0, \quad \int_{0}^{2\pi} d{\rm arg}(z')\ {z'}^{k} = 0, \quad (k \neq 0).
    \label{eq:orthogonality}
\end{align}
Thus, the normalization of localized modes can be determined in the following way:
\begin{align}
    1
    =& \int_{{\rm blow-up\ manifold}} dz d\bar{z} \sqrt{|{\rm det}(g)|} |\psi^{a_0}_{{\rm blow-up},0}|^2 \notag \\
    \simeq& \int_{r}^{\infty} d|z| |z| \int_{0}^{\frac{2\pi}{N}} d\varphi |C^{a_0}_{0}|^2 N^2 |z|^{2(m-(\ell_0-a_0)N)} e^{-\frac{\pi M}{{\rm Im}\tau}|z|^2} \notag \\
    &+ \int_{0}^{\frac{r}{N+1}} d|z'| |z'| \int_{0}^{2\pi} d\varphi \frac{4R^4}{\left( R^2 + |z'|^2 \right)^2} \frac{|{C'}^{a_0}_{0}|^2 |z'|^{2a_0}}{\left( R^2 + |z'|^2 \right)^{M'-1}} \notag \\
    \simeq& \left| C^{a_0}_{0} \right|^2 \pi \left(\frac{1}{r^2} \right)^{(\ell_0-a_0)N-(m+1)} \left[ N \frac{\left( -\frac{\pi M}{{\rm Im}\tau} r^2 \right)^{(\ell_0-a_0)N-(m+1)}}{[(\ell_0-a_0)N-(m+1)]!} E_1\left( \frac{\pi M}{{\rm Im}\tau}r^2 \right) +  L_0 \right],
    \label{eq:normallocal}
\end{align}
with 
\begin{align}
        L_0 \simeq& \left( \frac{N-1}{2N}(M'-a_0) \right)^{-1} \frac{1 - \sum_{p=0}^{a_0} \frac{M'!}{(M'-p)! p!} \left( \frac{N+1}{2N} \right)^{M'-p} \left( \frac{N-1}{2N} \right)^{p}}{\frac{M'!}{(M'-a_0)! a_0!} \left( \frac{N+1}{2N} \right)^{M'-a_0} \left( \frac{N-1}{2N} \right)^{a_0}} + \left( (\ell_0 - a_0) - \frac{m+1}{N} \right)^{-1}, \notag
\end{align}
where $E_1$ denotes the exponential integral.
The detailed calculation of Eq.~(\ref{eq:normallocal}) is shown in Appendix~\ref{ap:normallocal}.
Therefore, we obtained normalizable zero-mode wave functions in Eq.~(\ref{eq:wavlocalblowup}), and they correspond to localized modes under the orbifold limit $r \rightarrow 0$.
Similarly, the above analysis is valid for localized modes around other orbifold singular points by just replacement in Eq.~(\ref{eq:replace}).

% ------------------------------------------------------ %
% ------------------------------------------------------ %
% ------------------------------------------------------ %
% ------------------------------------------------------ %

\section{Yukawa couplings on magnetized blow-up manifolds of $T^2/Z_N$ orbifolds}
\label{sec:Yukawa}

Finally, we study Yukawa coupling of 4D effective theory derived from the magnetized blow-up manifold.
Here, we only replace the cone around $z_I=0$ with the part of $S^2$.
Similarly, we can consider the following analysis even at the other orbifold singular points.
First, we denote bulk zero-modes and localized zero-modes shortly as B and L, respectively.
When we consider the Yukawa coupling X$_1$-X$_2$-X$_3$ (X$=$B, L) in which $M_1+M_2=M_3$, $\xi^F_{01}+\xi^F_{02}=\xi^F_{03}$ ($\ell_{01}+\ell_{02}=\ell_{03}$, $m_1+m_2=m_3$\footnote{When $m_1+m_2=m_3+N$ and $\ell_{01}+\ell_{02}=\ell_{03}+1$ are satisfied, correction terms in Eq.~(\ref{eq:YukawaBBB}) are vanished,~i.e. $Y^{ijk}_{{\rm blow-up}}=Y^{ijk}_{T^2/Z_N}$, while they give corrections for B$_1$-L$_2$($b_0=\ell_{02}-1$)-B$_3$ coupling, $Y^{i(\ell_{02}-1)k}_{{\rm blow-up}}$, instead.}), and $(\alpha_1,\alpha_{\tau})_1+(\alpha_1,\alpha_{\tau})_2\equiv(\alpha_1,\alpha_{\tau})_3\ ({\rm mod}\ 1)$ are satisfied, only three patterns of couplings, (i) B$_1$-B$_2$-B$_3$ coupling, (ii) L$_1$-L$_2$-L$_3$ coupling, and (iii) B$_1$-L$_2$-L$_3$ coupling, are allowed by considering Eq.~(\ref{eq:orthogonality}).
Thus, we have a specific coupling selection rule in our theory.
We can calculate their Yukawa coupling by using the results of Eqs.~(\ref{eq:normalbulk}) and (\ref{eq:normallocal}).

In case (i), the Yukawa coupling in the 4D effective theory can be expressed as
\begin{align}
    Y^{ijk}_{{\rm blow-up}}
    =& y^{(3)}_{B_1-B_2-B_3}\int_{{\rm blow-up\ manifold}} dz d\bar{z} \sqrt{|{\rm det}(g)|} ( \psi^{k}_{{\rm blow-up},0} )^{\ast} \psi^{i}_{{\rm blow-up},0} \psi^{j}_{{\rm blow-up},0} \notag \\
    =& Y^{ijk}_{T^2/Z_N} - y^{(3)}_{B_1-B_2-B_3}\int_{0}^{r} d|z| |z| \int_{0}^{\frac{2\pi}{N}} d\varphi (C^{k}_{0})^{\ast} C^{i}_{0} C^{j}_{0} N^3 |z|^{2m_3} e^{-\frac{\pi M_3}{{\rm Im}\tau}|z|^2} \notag \\
    &+ y^{(3)}_{B_1-B_2-B_3}\int_{0}^{\frac{r}{N+1}} d|z'| |z'| \int_{0}^{2\pi} d\varphi \frac{4R^4}{\left( R^2 + |z'|^2 \right)^2} \frac{({C'}^{k}_{0})^{\ast} {C'}^{i}_{0} {C'}^{j}_{0} |z'|^{2\ell_{03}}}{\left( R^2 + |z'|^2 \right)^{M'_3-1}} \notag \\
    =& Y^{ijk}_{T^2/Z_N} +  y^{(3)}_{B_1-B_2-B_3}(C^{k}_{0})^{\ast} C^{i}_{0} C^{j}_{0} N \pi (r^2)^{m_3+1} B_{03},
    \label{eq:YukawaBBB} %\\
\end{align}
where $y^{(3)}_{B_1-B_2-B_3}$ denotes the 3-point coupling in higher dimensional theory, and 
$Y^{ijk}_{T^2/Z_N}$ denotes the 4D Yukawa coupling in the orbifold limit.
Note that we use wave functions in Eq.~(\ref{eq:wavbulk}).
When we calculate it by orthonormal basis, only $Y^{i'_{\text{max}}j'_{\text{max}}k'_{\text{max}}}_{{\rm blow-up}}$ receives the blow-up correction while the others remain $Y^{i'j'k'}_{{\rm blow-up}} = Y^{i'j'k'}_{T^2/Z_N}$.

In case (ii), the Yukawa coupling on the magnetized blow-up manifold can be expressed as
\begin{align}
    Y^{a_{0}b_{0}c_{0}}_{{\rm blow-up}}
    =& y^{(3)}_{L_1-L_2-L_3}\int_{{\rm blow-up\ manifold}} dz d\bar{z} \sqrt{|{\rm det}(g)|} ( \psi^{c_{0}}_{{\rm blow-up},0} )^{\ast} \psi^{a_{0}}_{{\rm blow-up},0} \psi^{b_{0}}_{{\rm blow-up},0} \notag \\
    \simeq& y^{(3)}_{L_1-L_2-L_3}\Biggl[ \int_{r}^{\infty} d|z| |z| \int_{0}^{\frac{2\pi}{N}} d\varphi (C^{c_{0}}_{0})^{\ast} C^{a_{0}}_{0} C^{b_{0}}_{0} N^3 |z|^{2(m_3-(\ell_{03}-c_{0})N)} e^{-\frac{\pi M_3}{{\rm Im}\tau}|z|^2} \notag \\
    &+ \int_{0}^{\frac{r}{N+1}} d|z'| |z'| \int_{0}^{2\pi} d\varphi \frac{4R^4}{\left( R^2 + |z'|^2 \right)^2} \frac{({C'}^{c_{0}}_{0})^{\ast} {C'}^{a_{0}}_{0} {C'}^{b_{0}}_{0} |z'|^{2c_{0}}}{\left( R^2 + |z'|^2 \right)^{M'_3-1}} \Biggl] \delta_{a_{0}+b_{0},c_{0}} \notag \\
    \simeq& y^{(3)}_{L_1-L_2-L_3}\frac{C^{a_{0}}_{0} C^{b_{0}}_{0}}{C^{c_{0}}_{0}} N \delta_{a_{0}+b_{0},c_{0}}, \label{eq:YukawaLLL}
\end{align}
where $y^{(3)}_{L_1-L_2-L_3}$ denotes the 3-point coupling in higher dimensional theory.

The case (iii) is the same as the case (ii) by replacing $a_{0}$ and $\delta_{a_0+b_0,c_0}$ with $i$ and $\delta_{\ell_{01}+b_0=c_0}$, respectively,~i.e.,
\begin{align}
    Y^{ib_{0}c_{0}}_{{\rm blow-up}} \simeq& y^{(3)}_{B_1-L_2-L_3}\frac{C^{i}_{0} C^{b_{0}}_{0}}{C^{c_{0}}_{0}} N \delta_{\ell_{01}+b_{0},c_{0}}, \label{eq:YukawaBLL}
\end{align}
where $y^{(3)}_{B_1-L_2-L_3}$ denotes the 3-point coupling in higher dimensional theory.

As a result, the Yukawa couplings among bulk modes (i) receive the contributions of blow-up radius, which play an important role in realizing the hierarchical structure of fermion masses as well as mixing angles, as demonstrated in Ref. \cite{Kobayashi:2019gyl}. By contrast, our results exhibit that Yukawa couplings including localized zero-modes are determined by the normalization factor depending on the localized flux. 
Similarly, we can compute higher dimensional operators.
The overall coefficients such as $y^{(3)}_{B_1-B_2-B_3}$, $y^{(3)}_{L_1-L_2-L_3}$, and $y^{(3)}_{B_1-L_2-L_3}$ depend on 
higher dimensional theory.
They may be unified in supersymmetric Yang-Mills theory on a smooth manifold.
All of the couplings originate from the gauge coupling in higher dimensional supersymmetric Yang-Mills theory, which is a
low-energy effective field theory of superstring theory.
Obviously, there is no difference between bulk and localized modes in a smooth manifold.
It is interesting to understand the flavor structure of localized modes as well as the origin of localized 
modes from the viewpoint of the string theory, but we leave the detailed study for future work.

% ------------------------------------------------------ %
% ------------------------------------------------------ %
% ------------------------------------------------------ %
% ------------------------------------------------------ %

\section{Conclusion}
\label{sec:conclusion}

We have studied the blow-up manifold of $T^2/Z_N$ orbifold with both bulk and localized magnetic fluxes.
On this background, we studied chiral zero-mode wave functions.
There are two types of matter zero-modes, namely bulk zero-modes and localized zero-modes.

For bulk zero-modes, although they have already been studied in Ref.~\cite{Kobayashi:2019fma}, we studied them again more precisely by introducing the singular gauge transformation to remove their $Z_N$ phase in their $Z_N$ twisted boundary condition.
Then, we can treat not only $Z_N$ invariant wave functions but also ones with $Z_N$ charge $m \neq 0$.
The normalization of zero-mode wave functions with arbitrary $Z_N$ charge was carefully calculated.

In addition, according to Ref.~\cite{Kobayashi}, localized flux induces new chiral zero modes. 
By explicitly computing their wave functions, we found that they correspond to localized zero modes at the orbifold singular point of $T^2/Z_N$ orbifold.
Although they diverge at the singular point, we calculated their normalization on the blow-up manifold to regularize them.
Moreover, by computing Yukawa coupling among bulk zero modes and localized zero modes, it turns out that only three patterns of Yukawa coupling are allowed. 
We have a specific coupling selection rule.
It would be interesting to study phenomenological implications of such coupling selection rule 
including higher dimensional operators.

It is interesting to apply our analysis for more general higher dimensional orbifolds such as $T^4/Z_N$ and $T^6/Z_N$ orbifolds
\footnote{The higher dimensional orbifold models with bulk magnetic fluxes were studied \cite{Abe:2014nla,Kikuchi:2022lfv}.}.
It is also important to study the relation with string theory.
For example, localized modes, i.e., twisted modes should appear massless in heterotic string theory on 
toroidal orbifold compactifications with generic gauge background by stringy consistency.
It would be important to revisit this aspect from the viewpoint of our analysis on localized gauge fluxes and localized modes.
However, that is beyond our scope and we would study elsewhere.

% ------------------------------------------------------ %
% ------------------------------------------------------ %

%-------- acknowledgement -------%

\vspace{1.5 cm}
\noindent
{\large\bf Acknowledgement}\\

This work was supported by JSPS KAKENHI Grants No. JP20K14477 (H. O.), JP 18K03649 (M.S.), JP 21J20739 (M. T.), and JP20J20388 (H. U.), and the Education and Research
Program for Mathematical and Data Science from the Kyushu University (H. O.). 
Y.T. is supported in part by Scuola Normale, by INFN (IS GSS-Pi) and by the MIUR-PRIN contract 2017CC72MK$\_$003.

% ------------------------------------------------------ %
% ------------------------------------------------------ %

%-------- Appendix -------%

\appendix

% ------------------------------------------------------ %
% ------------------------------------------------------ %
% ------------------------------------------------------ %
% ------------------------------------------------------ %

\section{Normalization of bulk zero modes}
\label{ap:normalbulk}

Here, we show the detailed calculation of Eq.~(\ref{eq:normalbulk}).
It consists of three terms.
The first term shows the calculation in all regions of the original $T^2/Z_N$ orbifold.
The second term shows the calculation in the region of the cone around $z_I=0$ which is cut out from the $T^2/Z_N$ orbifold.
The third term shows the calculation in the region of the part of $S^2$ which is embedded instead of the cone.
In the following, we show the detailed calculation of the second and third terms.

The second term can be calculated as
\begin{align}
    G_{ij}^{(2)}
    &\equiv \int_{0}^{r} d|z| |z| \int_{0}^{\frac{2\pi}{N}} d\varphi (C^{i}_{0})^{\ast} C^{j}_{0} N^2 |z|^{2m} e^{-\frac{\pi M}{{\rm Im}\tau}|z|^2} \notag \\
    &=  (C^{i}_{0})^{\ast} C^{j}_{0} \pi N \left(\frac{\pi M}{{\rm Im}\tau}\right)^{-(m+1)} \int_{0}^{\frac{\pi M}{{\rm Im}\tau}r^2} d\left( \frac{\pi M}{{\rm Im}\tau} |z|^2 \right) \left( \frac{\pi M}{{\rm Im}\tau} |z|^2 \right)^{m} e^{-\left(\frac{\pi M}{{\rm Im}\tau} |z|^2\right)} \notag \\
    &=  (C^{i}_{0})^{\ast} C^{j}_{0} \pi N \left(\frac{\pi M}{{\rm Im}\tau}\right)^{-(m+1)} \int_{0}^{\frac{\pi M}{{\rm Im}\tau}r^2} dt\ t^{(m+1)-1} e^{-t} \notag \\
    &=  (C^{i}_{0})^{\ast} C^{j}_{0} \pi N \left(\frac{\pi M}{{\rm Im}\tau}\right)^{-(m+1)} \gamma \left(m+1, \frac{\pi M}{{\rm Im}\tau}r^2 \right), \notag
\end{align}
where $\gamma(m+1, \frac{\pi M}{{\rm Im}\tau}r^2)$ denotes the lower incomplete gamma function.
It satisfies the following recurrence relation:
\begin{align}
    \begin{array}{l}
        \gamma \left(m+1, \frac{\pi M}{{\rm Im}\tau}r^2 \right) = m \gamma \left(m, \frac{\pi M}{{\rm Im}\tau}r^2 \right) -\left( \frac{\pi M}{{\rm Im}\tau}r^2 \right)^m e^{-\left( \frac{\pi M}{{\rm Im}\tau}r^2 \right)}  \\
        \gamma \left(1,\frac{\pi M}{{\rm Im}\tau}r^2 \right) = 1 - e^{-\left( \frac{\pi M}{{\rm Im}\tau}r^2 \right)} 
    \end{array}, \notag
\end{align}
and then by solving this recurrence relation, $\gamma(m+1, \frac{\pi M}{{\rm Im}\tau}r^2)$ can be expressed as
\begin{align}
    \gamma \left(m+1, \frac{\pi M}{{\rm Im}\tau}r^2 \right)
    &= m! e^{-\frac{\pi M}{{\rm Im}\tau}r^2} \left[ e^{\frac{\pi M}{{\rm Im}\tau}r^2}  - \sum_{p=0}^{m} \frac{1}{p!} \left( \frac{\pi M}{{\rm Im}\tau}r^2 \right)^{p} \right] \notag \\
    &= e^{-\frac{\pi M}{{\rm Im}\tau}r^2} \frac{1}{m+1} \left( \frac{\pi M}{{\rm Im}\tau}r^2 \right)^{m+1} \sum_{p=0}^{\infty} \frac{(m+1)!}{(m+1+p) !} \left( \frac{\pi M}{{\rm Im}\tau}r^2 \right)^{p}. \notag
\end{align}
Thus, the second term $G_{ij}^{(2)}$ can be expressed as
\begin{align}
    G_{ij}^{(2)} = (C^{i}_{0})^{\ast} C^{j}_{0} \pi (r^2)^{m+1} e^{-\frac{\pi M}{{\rm Im}\tau}r^2} \left( \frac{m+1}{N} \right)^{-1} \sum_{p=0}^{\infty} \frac{(m+1)!}{(m+1+p) !} \left( \frac{\pi M}{{\rm Im}\tau}r^2 \right)^{p}.
\end{align}
By contrast, the third term can be calculated as
\begin{align}
    G_{ij}^{(3)}
    \equiv& \int_{0}^{\frac{r}{N+1}} d|z'| |z'| \int_{0}^{2\pi} d\varphi \frac{4R^4}{\left( R^2 + |z'|^2 \right)^2} \frac{ ({C'}^{i}_{0})^{\ast} {C'}^{j}_{0} |z'|^{2\ell_0}}{\left( R^2 + |z'|^2 \right)^{M'-1}} \notag \\
    =&  ({C'}^{i}_{0})^{\ast} {C'}^{j}_{0} 4\pi (R^2)^{1-(M'-\ell_0-1)} \notag \\
    &\times \int_{\frac{N+1}{2N}}^{1} d\left( \frac{R^2}{R^2+|z'|^2} \right) \left( 1 - \frac{R^2}{R^2+|z'|^2} \right)^{\ell_0} \left( \frac{R^2}{R^2+|z'|^2} \right)^{M'-\ell_0-1} \notag \\
    =& (C^{i}_{0})^{\ast} C^{j}_{0} N^2 (r^2)^{m} e^{-\frac{\pi M}{{\rm Im}\tau}r^2} \left( \frac{r^2}{(N+1)^2} \right)^{M'-\ell_0-1} \left( \frac{2N}{N-1} \right)^{M'-1} 4\pi \left( \frac{r^2}{(N-1)(N+1)} \right)^{2+\ell_0-M'} \notag \\
    &\times \left( \int_{0}^{1} dt\ t^{(M'-\ell_0)-1} (1-t)^{(\ell_0+1)-1} - \int_{0}^{\frac{N+1}{2N}} dt\ t^{(M'-\ell_0)-1} (1-t)^{(\ell_0+1)-1} \right)  \notag \\
    =&  (C^{i}_{0})^{\ast} C^{j}_{0} \pi (r^2)^{m+1} e^{-\frac{\pi M}{{\rm Im}\tau}r^2} \left( \frac{2N}{N+1} \right)^{M'-\ell_0} \left( \frac{2N}{N-1} \right)^{\ell_0+1} \notag \\
    &\times \left( \beta(M'-\ell_0, \ell_0+1) - \beta_{\frac{N+1}{2N}}(M'-\ell_0,\ell_0+1) \right), \notag
\end{align}
where $\beta(M'-\ell_0, \ell_0+1)$ and $\beta_{\frac{N+1}{2N}}(M'-\ell_0,\ell_0+1)$ denote the beta function and the incomplete beta function, respectively.
They satisfy the following recurrence relations:
\begin{align}
    &\begin{array}{l}
        \beta(M'-\ell_0,\ell_0+1) = \frac{\ell_0}{M'-\ell_0} \beta_{\frac{N+1}{2N}}(M'-\ell_0+1,\ell_0) \\
        \beta(M',1) = \frac{1}{M'}
    \end{array}, \notag \\
    &\begin{array}{l}
        \beta_{\frac{N+1}{2N}}(M'-\ell_0,\ell_0+1) = \frac{1}{M'-\ell_0} \left( \ell_0 \beta_{\frac{N+1}{2N}}(M'-\ell_0+1,\ell_0) + \left( \frac{N+1}{2N} \right)^{M'-\ell_0} \left( \frac{N-1}{2N} \right)^{\ell_0} \right) \\
        \beta_{\frac{N+1}{2N}}(M',1) = \frac{1}{M'} \left( \frac{N+1}{2N} \right)^{M'}
    \end{array}, \notag
\end{align}
and then by solving these recurrence relations, they can be expressed as
\begin{align}
    \beta(M'-\ell_0,\ell_0-1) &= \frac{\Gamma(M'-\ell_0) \Gamma(\ell_0+1)}{\Gamma(M'+1)}, \notag \\
    \beta_{\frac{N+1}{2N}}(M'-\ell_0,\ell_0-1) &= \frac{\Gamma(M'-\ell_0) \Gamma(\ell_0+1)}{\Gamma(M'+1)} \sum_{p=0}^{\ell_0} \frac{\Gamma(M'+1)}{\Gamma(M'-p+1) \Gamma(p+1)} \left( \frac{N+1}{2N} \right)^{M'-p} \left( \frac{N-1}{2N} \right)^{p}, \notag
\end{align}
respectively.
Here, $\Gamma(X)$ denotes the gamma function, which satisfies the recurrence relation
\begin{align}
    \Gamma(X+1) = X\Gamma(X). \notag
\end{align}
Thus, the third term $G_{ij}^{(3)}$ can be expressed as
\begin{align}
    &G_{ij}^{(3)}
    = (C^{i}_{0})^{\ast} C^{j}_{0} \pi (r^2)^{m+1} e^{-\frac{\pi M}{{\rm Im}\tau}r^2} \left( \frac{N-1}{2N}(M'-\ell_0) \right)^{-1} \frac{1-\sum_{p=0}^{\ell_0} \frac{\Gamma(M'+1)}{\Gamma(M'-p+1) \Gamma(p+1)}  \left( \frac{N+1}{2N} \right)^{M'-p} \left( \frac{N-1}{2N} \right)^{p}}{\frac{\Gamma(M'+1)}{\Gamma(M'-\ell_0+1) \Gamma(\ell_0+1)}  \left( \frac{N+1}{2N} \right)^{M'-\ell_0} \left( \frac{N-1}{2N} \right)^{\ell_0}}.
\end{align}
By combining these results, we obtain Eq.~(\ref{eq:normalbulk}).

% ------------------------------------------------------ %
% ------------------------------------------------------ %
% ------------------------------------------------------ %
% ------------------------------------------------------ %

\section{Normalization of localized zero modes}
\label{ap:normallocal}

In this section, we show the detailed calculation of Eq.~(\ref{eq:normallocal}).
The first term shows the calculation in the bulk region, while the second term shows the calculation in the blow-up region.
The first term can be calculated as
\begin{align}
    & \int_{r}^{\infty} d|z| |z| \int_{0}^{\frac{2\pi}{N}} d\varphi |C^{a_0}_{0}|^2 N^2 |z|^{2(m-(\ell_0-a_0)N)} e^{-\frac{\pi M}{{\rm Im}\tau}|z|^2} \notag \\
    =& \left| C^{a_0}_{0} \right|^2 \pi N \left( \frac{\pi M}{{\rm Im}\tau} \right)^{(\ell_0-a_0)N-(m+1)} \int_{\frac{\pi M}{{\rm Im}\tau}r^2}^{\infty} d\left( \frac{\pi M}{{\rm Im}\tau}|z|^2 \right) \left( \frac{\pi M}{{\rm Im}\tau}|z|^2 \right)^{m-(\ell_0-a_0)N} e^{-\left( \frac{\pi M}{{\rm Im}\tau}|z|^2 \right)} \notag \\
    =& \left| C^{a_0}_{0} \right|^2 \pi N \left( \frac{\pi M}{{\rm Im}\tau} \right)^{(\ell_0-a_0)N-(m+1)} \int_{\frac{\pi M}{{\rm Im}\tau}r^2}^{\infty} dt\ t^{m-(\ell_0-a_0)N} e^{-t} \notag \\
    =& \left| C^{a_0}_{0} \right|^2 \pi N \left( \frac{\pi M}{{\rm Im}\tau} \right)^{(\ell_0-a_0)N-(m+1)} \Gamma \left( 1+m-(\ell_0-a_0)N, \frac{\pi M}{{\rm Im}\tau}r^2 \right), \notag
\end{align}
where $\Gamma \left( 1+m-(\ell_0-a_0)N, \frac{\pi M}{{\rm Im}\tau}r^2 \right)$ denotes the upper incomplete gamma function.
We note that $1+m-(\ell_0-a_0)N<0$.
Then, it satisfies the following recurrence relation:
\begin{align}
    \begin{array}{l}
        \Gamma \left( 1+m-(\ell_0-a_0)N, \frac{\pi M}{{\rm Im}\tau}r^2 \right) = \\
        \frac{1}{1+m-(\ell_0-a_0)N} \left( \Gamma \left( 2+m-(\ell_0-a_0)N, \frac{\pi M}{{\rm Im}\tau}r^2 \right) - \left( \frac{\pi M}{{\rm Im}\tau}r^2 \right)^{1+m-(\ell_0-a_0)N} e^{-\left( \frac{\pi M}{{\rm Im}\tau}r^2 \right)} \right) \\
        \Gamma \left( 0, \frac{\pi M}{{\rm Im}\tau}r^2 \right) = E_1\left( \frac{\pi M}{{\rm Im}\tau}r^2 \right) 
    \end{array}, \notag
\end{align}
where $E_1\left( \frac{\pi M}{{\rm Im}\tau}r^2 \right)$ denotes the exponential integral.
Note that if $\frac{\pi M}{{\rm Im}\tau}r^2$ is sufficiently large, the exponential integral obeys
\begin{align}
    E_1\left( \frac{\pi M}{{\rm Im}\tau}r^2 \right) \simeq e^{-\left( \frac{\pi M}{{\rm Im}\tau}r^2 \right)} \sum_{p=0} (-1)^{p} p! \left( \frac{\pi M}{{\rm Im}\tau}r^2 \right)^{-(p+1)}. \notag
\end{align}
By solving this recurrence relation, $\Gamma \left( 1+m-(\ell_0-a_0)N, \frac{\pi M}{{\rm Im}\tau}r^2 \right)$ can be expressed as
\begin{align}
    &\Gamma \left( 1+m-(\ell_0-a_0)N, \frac{\pi M}{{\rm Im}\tau}r^2 \right) \notag \\
    =& \frac{(-1)^{(\ell_0-a_0)N-(m+1)}}{[(\ell_0-a_0)N-(m+1)]!} \left[ E_1\left( \frac{\pi M}{{\rm Im}\tau}r^2 \right) - e^{-\left( \frac{\pi M}{{\rm Im}\tau}r^2 \right)} \sum_{p=0}^{(\ell_0-a_0)N-(m+2)} (-1)^{p} p! \left( \frac{\pi M}{{\rm Im}\tau}r^2 \right)^{-(p+1)}  \right]. \notag %\\
    %\simeq& \left( \frac{\pi M}{{\rm Im}\tau}r^2 \right)^{m-(\ell_0-a_0)N} e^{-\frac{\pi M}{{\rm Im}\tau}r^2} \sum_{p=0} \frac{[(\ell_0-a_0)N-(m+1)+p]!}{[(\ell_0-a_0)N-(m+1)]!} \left( - \frac{\pi M}{{\rm Im}\tau}r^2 \right)^{-p}. \notag
\end{align}
By contrast, the second term is the same as $G_{ij}^{(3)}$ in the previous section by replacing $\ell_0$ with $a_0$.
Thus, by combining these results, we obtain Eq.~(\ref{eq:normallocal}).

% ------------------------------------------------------ %
% ------------------------------------------------------ %
% ------------------------------------------------------ %
% ------------------------------------------------------ %


\begin{thebibliography}{99}

%%%%%%%%%%%%%%%% abstruct %%%%%%%%%%%%%%%%%%



%%%%%%%%%%%%%%%%% section 1 %%%%%%%%%%%%%%%%%

% --- blow-up and Calabi-Yau  ----- %

%\cite{Dixon:1985jw}
\bibitem{Dixon:1985jw}
L.~J.~Dixon, J.~A.~Harvey, C.~Vafa and E.~Witten,
%``Strings on Orbifolds,''
Nucl. Phys. B \textbf{261} (1985), 678-686
%doi:10.1016/0550-3213(85)90593-0


%\cite{Dixon:1986jc}
\bibitem{Dixon:1986jc}
L.~J.~Dixon, J.~A.~Harvey, C.~Vafa and E.~Witten,
%``Strings on Orbifolds. 2.,''
Nucl. Phys. B \textbf{274} (1986), 285-314
%doi:10.1016/0550-3213(86)90287-7




% --- wave of T2, T2/Z2, T2/ZN ----- %

%\cite{Cremades:2004wa}
\bibitem{Cremades:2004wa}
D.~Cremades, L.~E.~Ibanez and F.~Marchesano,
%``Computing Yukawa couplings from magnetized extra dimensions,''
JHEP \textbf{05} (2004), 079
%doi:10.1088/1126-6708/2004/05/079
[arXiv:hep-th/0404229 [hep-th]].
%298 citations counted in INSPIRE as of 31 Jul 2022

%\cite{Abe:2008fi}
\bibitem{Abe:2008fi}
H.~Abe, T.~Kobayashi and H.~Ohki,
%``Magnetized orbifold models,''
JHEP \textbf{09} (2008), 043
%doi:10.1088/1126-6708/2008/09/043
[arXiv:0806.4748 [hep-th]].
%89 citations counted in INSPIRE as of 31 Jul 2022

%\cite{Abe:2013bca}
\bibitem{Abe:2013bca}
T.~H.~Abe, Y.~Fujimoto, T.~Kobayashi, T.~Miura, K.~Nishiwaki and M.~Sakamoto,
%``$Z_N$ twisted orbifold models with magnetic flux,''
JHEP \textbf{01} (2014), 065
%doi:10.1007/JHEP01(2014)065
[arXiv:1309.4925 [hep-th]].
%58 citations counted in INSPIRE as of 31 Jul 2022


%%%%%%%%%%%%%%% three-generation models %%%%%%%%%%%%%%%%%%%%

%\cite{Abe:2008sx}
\bibitem{Abe:2008sx}
H.~Abe, K.~S.~Choi, T.~Kobayashi and H.~Ohki,
%``Three generation magnetized orbifold models,''
Nucl. Phys. B \textbf{814}, 265-292 (2009)
%doi:10.1016/j.nuclphysb.2009.02.002
[arXiv:0812.3534 [hep-th]].


%\cite{Abe:2015yva}
\bibitem{Abe:2015yva}
T.~h.~Abe, Y.~Fujimoto, T.~Kobayashi, T.~Miura, K.~Nishiwaki, M.~Sakamoto and Y.~Tatsuta,
%``Classification of three-generation models on magnetized orbifolds,''
Nucl. Phys. B \textbf{894}, 374-406 (2015)
%doi:10.1016/j.nuclphysb.2015.03.004
[arXiv:1501.02787 [hep-ph]].


%%%%%%%%%% index %%%%%%%%%%%%%%%%%

%\cite{Sakamoto:2020pev}
\bibitem{Sakamoto:2020pev}
M.~Sakamoto, M.~Takeuchi and Y.~Tatsuta,
%``Zero-mode counting formula and zeros in orbifold compactifications,''
Phys. Rev. D \textbf{102} (2020) no.2, 025008
%doi:10.1103/PhysRevD.102.025008
[arXiv:2004.05570 [hep-th]].

%\cite{Sakamoto:2020vdy}
\bibitem{Sakamoto:2020vdy}
M.~Sakamoto, M.~Takeuchi and Y.~Tatsuta,
%``Index theorem on $T^2/\mathbb{Z}_N$ orbifolds,''
Phys. Rev. D \textbf{103}, no.2, 025009 (2021)
%doi:10.1103/PhysRevD.103.025009
[arXiv:2010.14214 [hep-th]].


% --- Higher coupling  ----- %

%\cite{Abe:2009dr}
\bibitem{Abe:2009dr}
H.~Abe, K.~S.~Choi, T.~Kobayashi and H.~Ohki,
%``Higher Order Couplings in Magnetized Brane Models,''
JHEP \textbf{06} (2009), 080
%doi:10.1088/1126-6708/2009/06/080
[arXiv:0903.3800 [hep-th]].
%60 citations counted in INSPIRE as of 02 Aug 2022


%%%%%%%%%% Q and L masses and mixing %%%%%%%%%%%%%%%%%





%\cite{Abe:2012fj}
\bibitem{Abe:2012fj}
H.~Abe, T.~Kobayashi, H.~Ohki, A.~Oikawa and K.~Sumita,
%``Phenomenological aspects of 10D SYM theory with magnetized extra dimensions,''
Nucl. Phys. B \textbf{870}, 30-54 (2013)
%doi:10.1016/j.nuclphysb.2013.01.014
[arXiv:1211.4317 [hep-ph]].  

%\cite{Abe:2014vza}
\bibitem{Abe:2014vza}
H.~Abe, T.~Kobayashi, K.~Sumita and Y.~Tatsuta,
%``Gaussian Froggatt-Nielsen mechanism on magnetized orbifolds,''
Phys. Rev. D \textbf{90}, no.10, 105006 (2014)
%doi:10.1103/PhysRevD.90.105006
[arXiv:1405.5012 [hep-ph]].





%\cite{Fujimoto:2016zjs}
\bibitem{Fujimoto:2016zjs}
Y.~Fujimoto, T.~Kobayashi, K.~Nishiwaki, M.~Sakamoto and Y.~Tatsuta,
%``Comprehensive analysis of Yukawa hierarchies on $T^2/Z_N$ with magnetic fluxes,''
Phys. Rev. D \textbf{94}, no.3, 035031 (2016)
%doi:10.1103/PhysRevD.94.035031
[arXiv:1605.00140 [hep-ph]].



%\cite{Kobayashi:2016qag}
\bibitem{Kobayashi:2016qag}
T.~Kobayashi, K.~Nishiwaki and Y.~Tatsuta,
%``CP-violating phase on magnetized toroidal orbifolds,''
JHEP \textbf{04}, 080 (2017)
%doi:10.1007/JHEP04(2017)080
[arXiv:1609.08608 [hep-th]].




%\cite{Kikuchi:2021yog}
\bibitem{Kikuchi:2021yog}
S.~Kikuchi, T.~Kobayashi, Y.~Ogawa and H.~Uchida,
%``Yukawa textures in modular symmetric vacuum of magnetized orbifold models,''
PTEP \textbf{2022}, no.3, 033B10 (2022)
%doi:10.1093/ptep/ptac035
[arXiv:2112.01680 [hep-ph]].

%\cite{Kikuchi:2022geu}
\bibitem{Kikuchi:2022geu}
S.~Kikuchi, T.~Kobayashi, M.~Tanimoto and H.~Uchida,
%``Mass matrices with CP phase in modular flavor symmetry,''
[arXiv:2206.08538 [hep-ph]].
%3 citations counted in INSPIRE as of 18 Aug 2022

%\cite{Hoshiya:2022qvr}
\bibitem{Hoshiya:2022qvr}
K.~Hoshiya, S.~Kikuchi, T.~Kobayashi and H.~Uchida,
%``Quark and lepton flavor structure in magnetized orbifold models at residual modular symmetric points,''
[arXiv:2209.07249 [hep-ph]].


%%%%  modular symmetry %%%%%%%%%%%%%%%%%%%%

%\cite{Kobayashi:2018rad}
\bibitem{Kobayashi:2018rad} 
 T.~Kobayashi, S.~Nagamoto, S.~Takada, S.~Tamba and T.~H.~Tatsuishi,
 %``Modular symmetry and non-Abelian discrete flavor symmetries in string compactification,''
 Phys.\ Rev.\ D {\bf 97}, no. 11, 116002 (2018)
% doi:10.1103/PhysRevD.97.116002
 [arXiv:1804.06644 [hep-th]].
 %%CITATION = doi:10.1103/PhysRevD.97.116002;%% 

%\cite{Kobayashi:2018bff}
\bibitem{Kobayashi:2018bff}
T.~Kobayashi and S.~Tamba,
%``Modular forms of finite modular subgroups from magnetized D-brane models,''
Phys.\ Rev.\ D {\bf 99} (2019) no.4, 046001
%doi:10.1103/PhysRevD.99.046001
[arXiv:1811.11384 [hep-th]].

%\cite{Kariyazono:2019ehj}
\bibitem{Kariyazono:2019ehj}
Y.~Kariyazono, T.~Kobayashi, S.~Takada, S.~Tamba and H.~Uchida,
%``Modular symmetry anomaly in magnetic flux compactification,''
Phys. Rev. D \textbf{100}, no.4, 045014 (2019)
%doi:10.1103/PhysRevD.100.045014
[arXiv:1904.07546 [hep-th]].




%\cite{Ohki:2020bpo}
\bibitem{Ohki:2020bpo}
H.~Ohki, S.~Uemura and R.~Watanabe,
%``Modular flavor symmetry on a magnetized torus,''
Phys. Rev. D \textbf{102}, no.8, 085008 (2020)
%doi:10.1103/PhysRevD.102.085008
[arXiv:2003.04174 [hep-th]].



%\cite{Kikuchi:2020frp}
\bibitem{Kikuchi:2020frp}
S.~Kikuchi, T.~Kobayashi, S.~Takada, T.~H.~Tatsuishi and H.~Uchida,
%``Revisiting modular symmetry in magnetized torus and orbifold compactifications,''
Phys. Rev. D \textbf{102}, no.10, 105010 (2020)
%doi:10.1103/PhysRevD.102.105010
[arXiv:2005.12642 [hep-th]].




%\cite{Kikuchi:2020nxn}
\bibitem{Kikuchi:2020nxn}
S.~Kikuchi, T.~Kobayashi, H.~Otsuka, S.~Takada and H.~Uchida,
%``Modular symmetry by orbifolding magnetized $T^2\times T^2$: realization of double cover of $\Gamma_N$,''
JHEP \textbf{11} (2020), 101
%doi:10.1007/JHEP11(2020)101
[arXiv:2007.06188 [hep-th]].


%\cite{Kikuchi:2021ogn}
\bibitem{Kikuchi:2021ogn}
S.~Kikuchi, T.~Kobayashi and H.~Uchida,
%``Modular flavor symmetries of three-generation modes on magnetized toroidal orbifolds,''
Phys. Rev. D \textbf{104}, no.6, 065008 (2021)
%doi:10.1103/PhysRevD.104.065008
[arXiv:2101.00826 [hep-th]].

%\cite{Almumin:2021fbk}
\bibitem{Almumin:2021fbk}
Y.~Almumin, M.~C.~Chen, V.~Knapp-Perez, S.~Ramos-Sanchez, M.~Ratz and S.~Shukla,
%``Metaplectic Flavor Symmetries from Magnetized Tori,'' 
JHEP \textbf{05} (2021), 078
%doi:10.1007/JHEP05(2021)078
[arXiv:2102.11286 [hep-th]].

%\cite{Tatsuta:2021deu}
\bibitem{Tatsuta:2021deu}
Y.~Tatsuta,
%``Modular symmetry and zeros in magnetic compactifications,''
JHEP \textbf{10}, 054 (2021)
%doi:10.1007/JHEP10(2021)054
[arXiv:2104.03855 [hep-th]].


%\cite{Kikuchi:2022bkn}
\bibitem{Kikuchi:2022bkn}
S.~Kikuchi, T.~Kobayashi, K.~Nasu, H.~Uchida and S.~Uemura,
%``Modular symmetry anomaly and nonperturbative neutrino mass terms in magnetized orbifold models,''
Phys. Rev. D \textbf{105}, no.11, 116002 (2022)
%doi:10.1103/PhysRevD.105.116002
[arXiv:2202.05425 [hep-th]].



% --- phenomenological model of  T2/Z2, T2/ZN ----- %


% --- Calabi-Yau ----- %

%\cite{Ishiguro:2021drk}
\bibitem{Ishiguro:2021drk}
K.~Ishiguro, T.~Kobayashi and H.~Otsuka,
%``Hierarchical structure of physical Yukawa couplings from matter field K\"ahler metric,''
JHEP \textbf{07}, 064 (2021)
%doi:10.1007/JHEP07(2021)064
[arXiv:2103.10240 [hep-th]].

%\cite{Ishiguro:2021ccl}
\bibitem{Ishiguro:2021ccl}
K.~Ishiguro, T.~Kobayashi and H.~Otsuka,
%``Symplectic modular symmetry in heterotic string vacua: flavor, CP, and R-symmetries,''
JHEP \textbf{01}, 020 (2022)
%doi:10.1007/JHEP01(2022)020
[arXiv:2107.00487 [hep-th]].





% --- blow-up of CN/ZN ----- %

%\cite{GrootNibbelink:2007lua}
\bibitem{GrootNibbelink:2007lua}
S.~Groot Nibbelink, M.~Trapletti and M.~Walter,
%``Resolutions of C**n/Z(n) Orbifolds, their U(1) Bundles, and Applications to String Model Building,''
JHEP \textbf{03} (2007), 035
%doi:10.1088/1126-6708/2007/03/035
[arXiv:hep-th/0701227 [hep-th]].


%\cite{Leung:2019oln}
\bibitem{Leung:2019oln}
P.~Leung and H.~Otsuka,
%``Heterotic Stringy Corrections to Metrics of Toroidal Orbifolds and Their Resolutions,''
Phys. Rev. D \textbf{99} (2019) no.12, 126011
%doi:10.1103/PhysRevD.99.126011
[arXiv:1903.12144 [hep-th]].


%\cite{Eguchi:1978xp}
\bibitem{Eguchi:1978xp}
T.~Eguchi and A.~J.~Hanson,
%``Asymptotically Flat Selfdual Solutions to Euclidean Gravity,''
Phys. Lett. B \textbf{74} (1978), 249-251.
%doi:10.1016/0370-2693(78)90566-X


% --- blow-up of T2/ZN ----- %

%\cite{Kobayashi:2019fma}
\bibitem{Kobayashi:2019fma}
T.~Kobayashi, H.~Otsuka and H.~Uchida,
%``Wavefunctions and Yukawa couplings on resolutions of T$^{2}$/\ensuremath{\mathbb{Z}}$_{N}$ orbifolds,''
JHEP \textbf{08} (2019), 046
%doi:10.1007/JHEP08(2019)046
[arXiv:1904.02867 [hep-th]].


%\cite{Kobayashi:2019gyl}
\bibitem{Kobayashi:2019gyl}
T.~Kobayashi, H.~Otsuka and H.~Uchida,
%``Flavor structure of magnetized $T^2/\mathbb{Z}_2$ blow-up models,''
JHEP \textbf{03} (2020), 042
%doi:10.1007/JHEP03(2020)042
[arXiv:1911.01930 [hep-ph]].


% --- localized mode of T2/ZN ----- %

%\cite{Lee:2003mc}
\bibitem{Lee:2003mc}
H.~M.~Lee, H.~P.~Nilles and M.~Zucker,
%``Spontaneous localization of bulk fields: The Six-dimensional case,''
Nucl. Phys. B \textbf{680} (2004), 177-198
%doi:10.1016/j.nuclphysb.2003.12.031
[arXiv:hep-th/0309195 [hep-th]].


%\cite{Abe:2020vmv}
\bibitem{Abe:2020vmv}
H.~Abe, T.~Kobayashi, S.~Uemura and J.~Yamamoto,
%``Loop Fayet-Iliopoulos terms in $T^2/Z_2$ models: Instability and moduli stabilization,''
Phys. Rev. D \textbf{102}, no.4, 045005 (2020)
%doi:10.1103/PhysRevD.102.045005
[arXiv:2003.03512 [hep-th]].
%\cite{Kobayashi}


\bibitem{Kobayashi}
T.~Kobayashi, H.~Otsuka, M.~Sakamoto, M.~Takeuchi, Y.~Tatsuta and H. Uchida,
``Index theorem on magnetized blow-up manifold of $T^2/\mathbb{Z}_N$".

%%%%%%%%%%%%%%%%% section 3 %%%%%%%%%%%%%%%%%

% --- wave of S2 ----- %

%\cite{Conlon:2008qi}
\bibitem{Conlon:2008qi}
J.~P.~Conlon, A.~Maharana and F.~Quevedo,
%``Wave Functions and Yukawa Couplings in Local String Compactifications,''
JHEP \textbf{09} (2008), 104
%doi:10.1088/1126-6708/2008/09/104
[arXiv:0807.0789 [hep-th]].


%\cite{Dolan:2020sjq}
\bibitem{Dolan:2020sjq}
B.~P.~Dolan and A.~Hunter-McCabe,
%``Ground state wave functions for the quantum Hall effect on a sphere and the Atiyah-Singer index theorem,''
J. Phys. A \textbf{53} (2020) no.21, 215306
%doi:10.1088/1751-8121/ab85e1
[arXiv:2001.02208 [hep-th]].


% --- + localized mode of T2/ZN ----- %

%\cite{Polchinski}
\bibitem{Polchinski}
J. Polchinski, “String Theory. Vol. 1: An Introduction To The Bosonic String,”, section 7.2.


%%%%%%%%% T^4/ZN %%%%%%%%%%%%%%%%%%

%\cite{Abe:2014nla}
\bibitem{Abe:2014nla}
H.~Abe, T.~Kobayashi, H.~Ohki, K.~Sumita and Y.~Tatsuta,
%``Non-Abelian discrete flavor symmetries of 10D SYM theory with magnetized extra dimensions,''
JHEP \textbf{06}, 017 (2014)
%doi:10.1007/JHEP06(2014)017
[arXiv:1404.0137 [hep-th]].


%\cite{Kikuchi:2022lfv}
\bibitem{Kikuchi:2022lfv}
S.~Kikuchi, T.~Kobayashi, K.~Nasu and H.~Uchida,
%``Classifications of magnetized T$^{4}$ and T$^{4}$/Z$_{2}$ orbifold models,''
JHEP \textbf{08}, 256 (2022)
%doi:10.1007/JHEP08(2022)256
[arXiv:2203.01649 [hep-th]].

%%%%%%%%%%%%%%%%%%%%%%%%%%%%%%%%%%%%%%%%%%%%%

\end{thebibliography}
\end{document}